\documentclass[onecolumn,amsmath,amssymb,superscriptaddress,nofootinbib,11pt,a4paper]{revtex4}

\pdfoutput=1
\usepackage[T1]{fontenc}
\usepackage{CJK}
\usepackage{xcolor}
\usepackage{amsfonts}
\usepackage{epstopdf}
\usepackage{wrapfig} 
\usepackage{subfigure}
\usepackage{graphicx}  
\usepackage{dcolumn}   
\usepackage{bm}
\usepackage{float}
\usepackage[T1]{fontenc}
\usepackage{CJK}
\usepackage{xcolor}
\usepackage{amsfonts}
\usepackage{epstopdf}
\usepackage{wrapfig} 
\usepackage{subfigure}
\usepackage{graphicx}  
\usepackage{dcolumn}   
\usepackage{bm}
\usepackage{float}
\newcommand{\be}{\begin{equation}}
\newcommand{\ee}{\end{equation}}
 \newcommand{\bea}{\begin{eqnarray}}
 \newcommand{\ena}{\end{eqnarray}}

\newcommand{\bra}{\langle}
\newcommand{\ket}{\rangle}

\newcommand{\when}{\bigg\vert}

\newcommand{\notice}{\equiv}

\newcommand{\exd}{\mathrm{d}}

\newcommand{\pd}{\partial}

\newcommand{\diff}[2]{\frac{\exd{#1}}{\exd{#2}}}

\usepackage{amsthm,amsmath,amssymb}
\usepackage{mathrsfs}

\begin{document}
\title{Holographic Einstein rings of Non-commutative black holes}
\author{Xin-Yun Hu}
\affiliation{College of Economic and Management, Chongqing Jiaotong University, Chongqing 400074, China}
\author{Xiao-Xiong Zeng}\email{xxzengphysics@163.com}
\affiliation{Department of Mechanics, Chongqing Jiaotong University, Chongqing 400074, China}
\author{Li-Fang Li}{}\thanks{Corresponding author: lilifang@imech.ac.cn}
\affiliation{Center for Gravitational Wave Experiment, National Microgravity Laboratory, Institute of Mechanics, Chinese Academy of Sciences, Beijing 100190, China.}
\author{Peng Xu}
\affiliation{Center for Gravitational Wave Experiment, National Microgravity Laboratory, Institute of Mechanics, Chinese Academy of Sciences, Beijing 100190, China.}
\affiliation{Lanzhou Center of Theoretical Physics, Lanzhou University, No. 222 South Tianshui Road, Lanzhou 730000, China}

\begin{abstract}
{With the help of AdS/CFT correspondence, we derive the desired response function of QFT on the boundary of the non-commutative black hole. Using the virtual optical system with a convex lens, we  obtain the Einstein rings of the black hole from the response function. All the results show that the holographic ring always appears with the concentric stripe when the observer located at the north pole. With the change of the observation position, the ring   changes into a luminosity-deformed ring, or bright spot.
 We also investigate the effect of the  non-commutative 
parameter on the ring and find   the ring radius becomes  larger as  the parameter increases. The effect of the temperature on the ring radius  is also investigated,  it is found that the  higher the temperature, the smaller the  ring radius. 
In addition, we also obtain the ingoing angle of the photon via geometric optics, as expected, this angle is consistent well with  the angle of the Einstein ring obtained via holography.  

}
\end{abstract}

\maketitle
 \newpage
 
\section{Introduction}
In recent years, non-commutative spacetime in gravity theories has become an important research subject~\cite{Nicolini09} in that it is considered as an alternative way to the quantum gravity~\cite{Snyder1947}. Many investigations in non-commutative gravity have been done, please see the comprehensive reviews~\cite{Szabo2006}. In particular, the effects of non-commutativity on black hole physics have attracted much attention, mainly because  that the final stage of the noncommutative  black hole is more abundant. As is well known, the non-commutativity eliminates point-like structures in favor of smeared objects in flat spacetime~\cite{Smailagic2006,Smailagic2003} and can be implemented in General Relativity by modifying the matter source~\cite{Nicolini2006}. Therefore, non-commutativity is introduced by modifying mass density so that the Dirac delta function is replaced by a Gaussian distribution~\cite{Nicolini2006} or alternatively by a Lorentzian distribution~\cite{Nozari2008,Anacleto2020}. In this way the mass density takes the form $\rho_{n}(r)=\frac{M\sqrt{n}}{\pi^{3/2}(r^2+\pi n)^2}$, where $n$ is the noncommutative parameter and $M$ is the total mass diffused throughout the region of linear size $n$. With this model in hand, we aim to analyze the lensed effect of a noncommutative black hole in the holographic framework closely followed~\cite{Hashimoto:2018okj,Hashimoto:2019jmw}.

In the paper~\cite{
Hashimoto:2018okj,Hashimoto:2019jmw}, they proposed a direct procedure to construct holographic images of the black hole in the
bulk from a given response function of the QFT on the boundary with the AdS/CFT correspondence. In the holographic picture, the response function with respect to an external source corresponds to the asymptotic data of
the bulk field generated by the source on the AdS boundary. For a thermal state on two-dimensional sphere dual to Schwarzschild AdS$_4$ black hole, they demonstrated that the holographic images gravitationally lensed
by the black hole can be constructed from the response function. And all these results are consistent with the
size of the photon sphere of the black hole calculated by geometrical optics. Closely followed by these breakthroughs, the authors in~\cite{Liu:2022cev,Zeng:2023zlf,Zeng:2023tjb,Hu:2023eoa, CEJM,Hu:2023mai} showed that this holographic images do exist in different gravitational backgrounds. However, the photon sphere varies according to the specific bulk dual geometry and the detailed behavior of Einstein ring also varies. Therefore, in this paper, we are tempted to investigate the behavior of the lensed response for the noncommutative black hole and study the effect of the noncommutative parameters on the lensed response.  

This paper is arranged as follows. In section~\ref{sec1}, we briefly review the noncommutative solution in spherically symmeric AdS black hole and introduce the optical system used to observe the Einstein rings. In section ~\ref{sec2}, we first give an explicit constrain on the non-commutative parameter $n$. And then we set up a holographic Einstein ring model and analyze the lensed response function. With the above optical device, we observe the Einstein ring in our model. Section~\ref{sec4} is the comparison between the holographic method and the geometrical optics method. Our result shows that the position of photon ring obtained from the geometrical optics is full consistence with that of the holographic method. Section~\ref{sec5} is devoted to our conclusions.

\section{Reivew of the holographic construction of Einstein ring in AdS black holes}\label{sec1}
Gravitational lensing is one of the fundamental phenomena caused  by strong
gravity.
Supposing there is a light source behind a gravitational
body, the observers will see a ring-like image of the light source, i.e., the so-called Einstein ring when the light source, the gravitational body, and observers are in
alignment.
If the gravitational body is a black hole, some light rays are so strongly
bended that they can go around the black hole many times, and especially
infinite times on the photon sphere.
As a result, multiple Einstein rings which correspond to winding numbers of
the light ray orbits emerge and infinitely concentrate
on the photon sphere. Recently, an observational
project for imaging black holes which is called the Event Horizon Telescope (EHT), has captured the first image of the
supermassive black hole in M87$^{\star}$ and   Sagittarius A$^{\star}$ ~\cite{EHT,EventHorizonTelescope:2022wkp}. And the dark area inside the photon sphere is named black hole shadow~\cite{Falcke:1999pj} and the shadow of a black hole contains a lot of information. The study of shadow not only enables us to comprehend the geometric structure of spacetime, but also helps us to explore various gravity models more deeply. Following~\cite{
Hashimoto:2018okj,Hashimoto:2019jmw}, the holographic image of an AdS black hole in the bulk was constructed when the wave emitted by the source at the boundary of AdS enters the bulk and then propagates in the bulk by considering the AdS/CFT correspondence~\cite{Hashimoto:2018okj,Hashimoto:2019jmw,Liu:2022cev}. Here we first review explicitly the construction of holographic ``images'' of the dual black hole
from the response function of the boundary QFT with external sources.

\begin{figure}[ht]
	\centering
\includegraphics[trim=0.4cm 0.4cm 1.4cm 0.1cm, clip=true, scale=0.8]{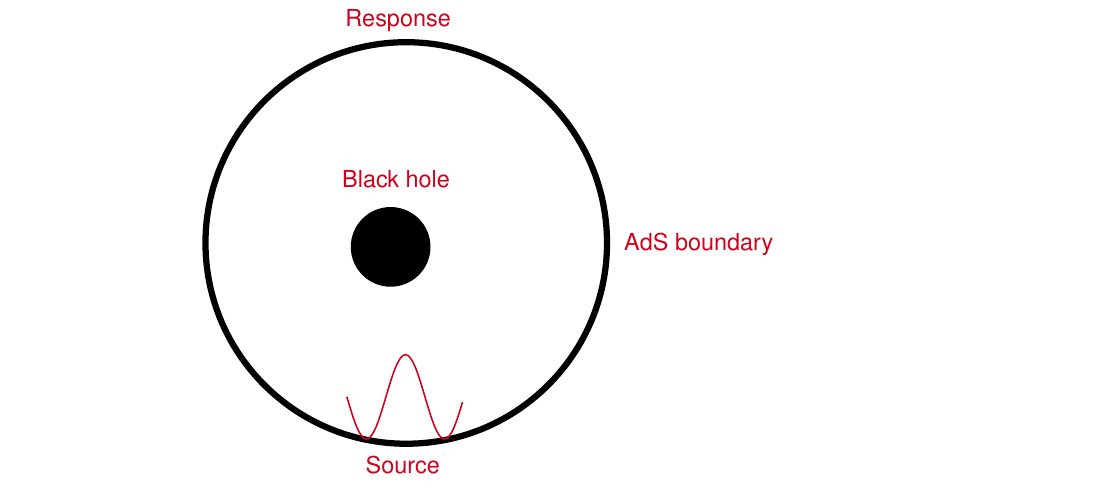}
\caption{A monochromatic Gaussian source $J_{\cal O}$ is located at a point on the AdS boundary. Its response is observed at another point on the same boundary\text{}.}\label{source_and_response}
\end{figure}

\begin{figure}[ht]
	\centering
 \includegraphics[height=2.5in]{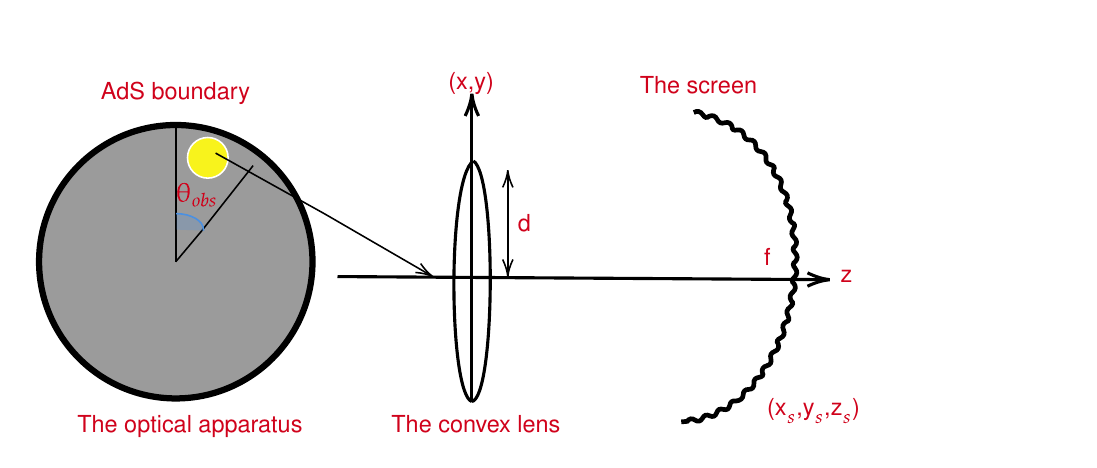}
	\caption{The observer and its \text{}telescope. }\label{optics}
\end{figure}

Considering a $(2+1)$-dimensional
boundary conformal field theory on a 2-sphere $S^2$ at a finite temperature $T$,
we study
a one-point function of a scalar operator ${\cal O}$ with its conformal dimension $\Delta_{\cal O} = 3$,
under a time-dependent localized source $J_{\cal O}$.
The schematic picture of our setup is shown in Fig.~\ref{optics}.
For the source $J_{\cal O}$,   we employ a time-periodic localized Gaussian source with the frequency $\omega$, 
amounts to an AdS boundary condition for the scalar field~(see Fig.~\ref{source_and_response}). The simplest form of the source $J_{\cal O}$ is chosen a monochromatic and axisymmetric Gaussian wave packet centered on the south pole $\theta_S=\pi$. Therefore, the expression of $J_{\cal O}$ in the ingoing Eddington coordinate is as follows 
\begin{eqnarray} \label{source}
\text{}\text{} J_\mathcal{O}(v_e,\theta)\text{}&=&\frac{1}{2\pi \text{}\delta^2}e^{-i\omega v_e}\exp\left[-\frac{(\pi-\theta)^2}{2\delta^2}\right]\text{}\nonumber\\
&=&e^{-i\omega {v_e}}\sum_{l=0}^\infty {c}_{l0}Y_{l0}(\theta),
\end{eqnarray}
where $v_e$ is the ingoing Eddington coordinate, $\delta$  represents the width of the wave produced by the Gaussian source $J_{\cal O}$ and $Y_{l0}$ is the spherical harmonics function.
We set the wave packet size $\delta$ to be $\delta\ll\text{}\pi$, and the coefficients of the spherical harmonics $Y_{l0}(\theta)$ is 
\begin{equation}
 	\text{}c_{l0}=(-1)^l(\frac{l+1/2}{2\pi})^{1/2}\exp\text{}\left[-\text{}\frac{ (l+1/2)^2\delta^2}{2}\right].
  \label{coefficient}
\end{equation}

  The time-dependent  scalar wave propagates inside the black hole spacetime
and reaches other points on the $S^2$ of the AdS boundary (please see Fig.~\ref{source_and_response}).
Using the optical system shown in Fig.~\ref{optics}, we are able to measure the local response function $\text{}e^{-i\omega t}\langle {\text{}\cal O}(\vec{x})\rangle\text{}$ which
contains the information about the bulk geometry of the black hole spacetime~\cite{Sharma2006}. Explicitly, such optical system consists of a convex lens and a spherical screen. In Fig.~\ref{optics}, the middle position is the lens with focal length $f$ regarded as a “converter” between
plane and spherical waves and located at $z=0$.  Consider that a plane wave is irradiated to the lens from the left hand side shown in Fig.~\ref{optics}. Such plane wave is converted into spherical wave and converges at the
focus $\text{}z = f$. We denote $\text{}\Psi_I$ and $\Psi_T$ as the incident wave and the transmitted wave. The conversion between $\Psi_I$ and $\Psi_T$ with frequency $\omega$ on the lens can be mathematically expressed as
\begin{equation}
\text{}\Psi_T(\vec{x})=e^{-i\frac{\omega|\vec{x}|^2} {2f}}\Psi_I(\vec{x})\text{}.
\label{psi_1}
\end{equation}

We consider a spherical screen located at $\text{}(x,y,z)=\text{}(x_{s},y_{s},z_{s})$ with $\text{}x_{s}^2+y_{s}^2+z_{s}^2=\text{}f^2$. The transmitted wave $\Psi_T$ converted by the lens is focusing and imaging on this screen. The wave function $\Psi_{s}(\vec{x_{s}})$ on the screen is given by
\begin{equation}
\text{}\text{}\Psi_{s} (\vec{x}_{s})=\text{}\int_{|\vec{x}|\leq d} \text{}\Psi_T (\vec{x})e^{i\omega L}d^2 x\text{},
\label{equation_5}
\end{equation} 
where $\text{}L$ is the distance between $\text{}(x,y,0)$ on the lens and $(x_{s},y_{s},z_{s})$ on the screen. Substituting Eq.(\ref{psi_1}) into Eq.(\ref{equation_5}), we have
\begin{equation}
\text{}\Psi_{s}(\text{}\vec{x}_{s})=\int_{|\vec{x}|\leq d}\text{} \Psi_I(\vec{x})e^{-i\frac{\omega}{f}\vec{x}\cdot\vec{x}_{s} }\text{}d^2 x \text{}.
\label{image_equation}
\end{equation}

Using a wave-optical method,
we have a formula which helps us convert the response function $\langle {\cal O}(\vec{x})\rangle\text{}$ to the image
of the dual black hole $|\Psi_\mathrm{s}(\vec{x}_\mathrm{s})|^2$ on a virtual screen shown as follows
\begin{equation}
 \text{}\Psi_\mathrm{s}(\vec{x}_\mathrm{s})=\text{} \int_{|\vec{x}|< d}\, \langle {\cal O}(\vec{x})\rangle e^{-i\frac{\omega}{f}\vec{x}\cdot\vec{x}_\mathrm{s}}d^2x\text{}\ ,
\label{transformation}
\end{equation}
here $\text{}\vec{x}=(x,y)$ is Cartesian-like coordinates on boundary and $\text{}\vec{x}_{s}=(x_{s},y_{s})$ is Cartesian-like coordinates on
the virtual screen. Next, we will employ  Eq.~\ref{transformation} 
to obtain the Einstein ring.  

\section{The holographic setup of non-commutative black holes}
\label{sec2}

\begin{figure}[ht]
	\centering
	\includegraphics[height=2.2in]{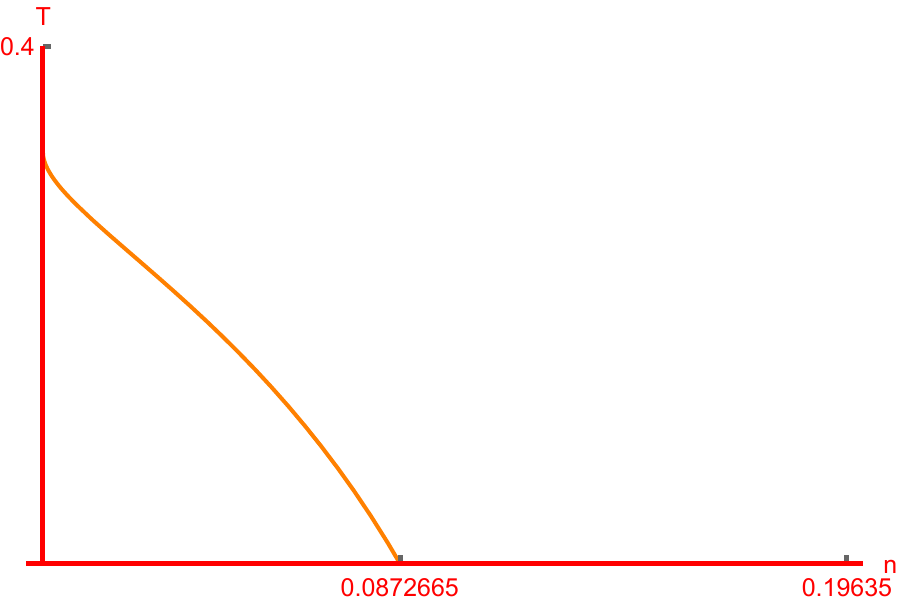}
	\caption{The relation between the temperature $T$ and the noncommutative parameter $n$  for the case $z_h=1$.}\label{temperature_1}
\end{figure}

\begin{figure}[ht]
	\centering
	\includegraphics[height=2.2in]
 {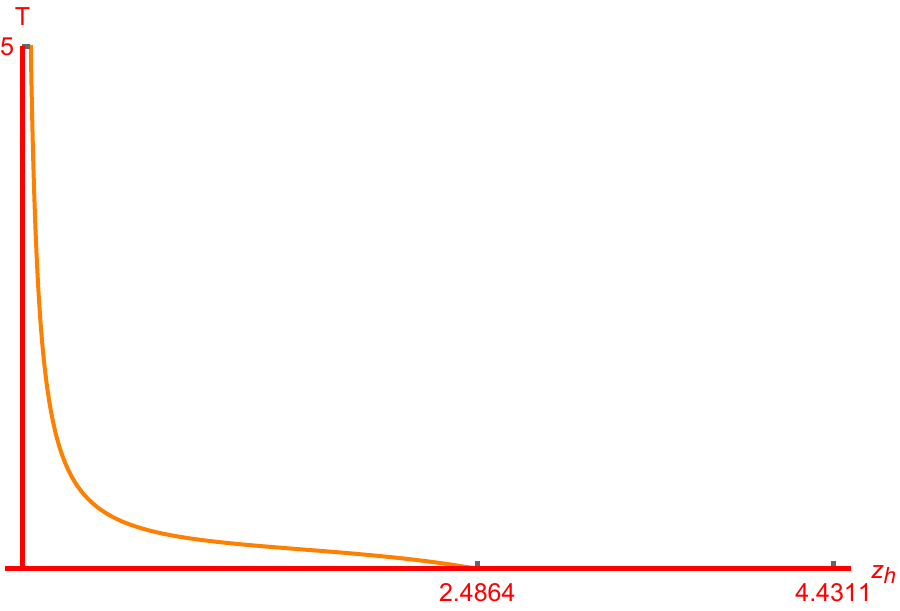}
	\caption{The relation between the temperature $T$ and $z_h$  for  a fixed noncommutative parameter $n=0.01$.}\label{temperature_2}
\end{figure}

In this section, we construct the holographic model for the non-commutative Schwarzschild black hole and study the properties of the response function carefully which further helps us to study the Einstein ring. The mass density of a static, spherically symmetric, particle-like gravitational source is no longer a function distribution, but given by a Lorentzian distribution shown as ~\cite{Nicolini2006, Nozari2008,Anacleto2020}
\begin{equation}
 \text{}\rho _{n}=\frac{\sqrt{n} M}{\pi ^{3/2} \text{}\left(\pi n+\text{}r^2\right)^2}\label{EQ2.1},
\end{equation}
here $n$ is the strength of non-commutativity of spacetime and $M$ is the total mass diffused throughout a region with linear size $\sqrt{n}$. For the smeared matter distribution, we further obtain~\cite{Anacleto2020}
\begin{equation}
\text{}\mathcal{M}_{n}=\int_0^r \rho _{n} (r) 4 \pi  r^2  dr=\text{}\frac{2 M}{\pi }\left(\tan^{-1}(\frac{r}{\sqrt{\pi n}})-\frac{\sqrt{\pi n} r}{\pi  n+r^2}\right)=\text{}-\frac{4 \sqrt{n} M}{\sqrt{\pi } r}+M+ \text{}\mathcal{O}(n^{3/2}). \label{EQ2.2}
\end{equation}
In this case, the non-commutative Schwarzschild black hole metric is given by
\begin{equation}
\text{}d s^2=\text{}-F(r) d t^2+\frac{1}{F (r)}d r^2+r^2\text{} \left(d \theta ^2+\sin ^2  \text{}\theta d \xi^2   \right)\text{},\label{EQ2.3}
\end{equation}
with
\begin{equation}
\text{}F (r)=1-\frac{2 \mathcal{M}_{n}}{r}+\frac{r^2}{l^2}=1-\frac{2 M}{r}+\frac{8 \sqrt{n} M}{\sqrt{\pi } r^2}+\frac{r^2}{l^2}+\mathcal{O}(n^{3/2})\text{}.\label{EQ2.4}
\end{equation}
here $l$ is the AdS radius, and we set $\text{}l=1$ in the following. According to  Eq.(\ref{EQ2.4}), we can express the temperature as the inverse of the event horizon 
\begin{equation}
\text{}M=\frac{\sqrt{\pi } \left(\frac{1}{z_h^2}+1\right)}{2 z_h^2 \left(\frac{\sqrt{\pi }}{z_h}-4 \sqrt{n}\right)},\label{mass}
\end{equation}
in which $z_h=1/r_h$. The mass should be non-vanishing, which leads to the following constraint relations  
\begin{equation}
n<\frac{\pi }{16 z_h^2} ~ \text{or}    ~
 z_h<\frac{\sqrt{\pi }}{4 \sqrt{n}}. \label{constrain}
\end{equation}
At the event horizon, the corresponding Hawking temperature is
\begin{equation}
T=\frac{\sqrt{\pi } \left(z_h^2+3\right)-8 \sqrt{n} z_h \left(z_h^2+2\right)}{4 \pi z_h \left(\sqrt{\pi }-4 \sqrt{n} z_h\right)},\label{temperature}
\end{equation}
The denominator in  Eq.(\ref{temperature}) is the same as that in Eq.(\ref{mass}). To assure the  temperature is positive, we also should impose the  numerator  in  Eq.(\ref{temperature})  is positive, which means

\begin{equation}
n<
\frac{\pi  \left(\text{zh}^4+6 \text{zh}^2+9\right)}{64 \text{zh}^2 \left(\text{zh}^2+2\right)^2} ~ \text{or}~  
z_h<-\frac{\sqrt[3]{K}}{24 \sqrt{n}}+\frac{384 n-\pi }{24 \sqrt[3]{K} \sqrt{n}}+\frac{\sqrt{\pi }}{24 \sqrt{n}}, \label{constrain2}
\end{equation}
in which 
\begin{equation}
 K=\sqrt{56623104 n^3+3621888 \pi  n^2+5184 \pi ^2 n}-2016 \sqrt{\pi } n-\pi ^{3/2}. 
\end{equation}
 In fact, the constraint in Eq.(\ref{constrain2}) is much stricter than that in Eq.(\ref{constrain}). For a fixed the parameter $z_h=1$, from  Eq.(\ref{constrain2}), we know $n$ should be smaller than 0.19635, but from Eq.(\ref{constrain}), we know $n$ should be smaller than 0.0872665, please see Fig.~\ref{temperature_1}, which describes the temperature $T$ decreases monotonically as the non-commutative parameter $n$ increases. For a fixed $n$, we also can find the relation  between $T$
 and $z_h$, which is shown in  Fig.~\ref{temperature_2}. We find Eq.(\ref{constrain2}) will produce a constraint $z_h<4.4311$  while Eq.(\ref{constrain}) produces  $z_h<2.4864$. In Fig.~\ref{temperature_2}, we  also see the temperature $T$ decreases monotonically 
 as the event horizon $z_h$ increases. These constraints are very important for our later numerical simulation because only parameters that are physically meaningful are important.

We take the complex scalar field as a probe field in the above non-commutative Schwarzschild background. The corresponding dynamics is determined by the Klein-Gordon equation
\begin{equation} \label{EOM}
	\text{}D_b D^b\Phi-\mathcal{M}^2 \Phi=0\text{},
\end{equation}
here $D_a=\nabla_a$ is the covariant operator and $\Phi$ is a complex scalar field with $\mathcal{M}$ its mass. And we take $\mathcal{M}^2=-2$ in the following. 

As before, we prefer the ingoing Eddington coordinate in order to solve the above Eq.~(\ref{EOM}) in a more convenient way, that is,
\begin{equation}
	\text{}v_e= t+z_*=t-\int \frac{1}{F(z)}\exd z\text{},
\end{equation}
here $z=1/r$ and 
\begin{equation} \label{fz}
\text{} F(z)=\frac{8 M \sqrt{n} z^4}{\sqrt{\pi }}-2 M z^3+z^2+1\text{}.
\end{equation}
Therefore the non-vanishing bulk background fields are transformed into the following smooth form
\begin{eqnarray}
	\text{}&d s^2=\frac{1}{z^2}\left[-F(z)d v_e^2-2 d z d v_e+d\Omega^2\right]\text{}. \label{newmetric1}
\end{eqnarray}
With $\text{}\Phi=z\mathscr{\phi}\text{}$, the asymptotic behaviour of $\text{}\mathscr{\phi}$ near the AdS boundary is expressed as
\begin{equation}
\text{}\mathscr{\phi}(v_e,z,\theta,\text{}\xi)=J_\mathcal{O}(v_e,\theta,\xi)
	+ z\bra\mathcal{O}\ket +\text{}O(z^2)\text{}.
\end{equation}
In the holographic dictionary, $\text{}J_\mathcal{O}$ is regarded as the source for the boundary field theory. And the response function, which is the corresponding expectation value of the dual operator, is given by
\begin{align}\text{}\bra\mathcal{O}\ket_{J_\mathcal{O}}=\text{}\bra\mathcal{O}\ket-\pd_{v_e} J_\mathcal{O}\text{},
\end{align}
where $\text{}\bra\mathcal{O}\ket$ obviously corresponds to the expectation value of the dual operator with the source turned off.

\begin{figure}
    \centering
    \subfigure {
        \includegraphics[width=3in]{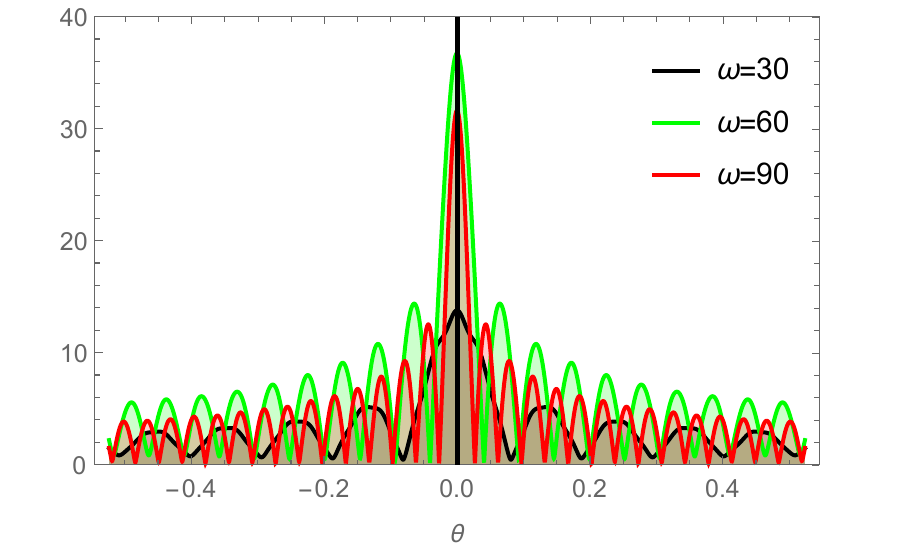}
    }
   \caption{The amplitude of $\bra \mathcal{O}\ket$ for different $\omega$  with  $z_h=1$ and $n=0.04$.}\label{omega}
\end{figure}

\begin{figure}
    \centering\subfigure{
        \includegraphics[width=3in]{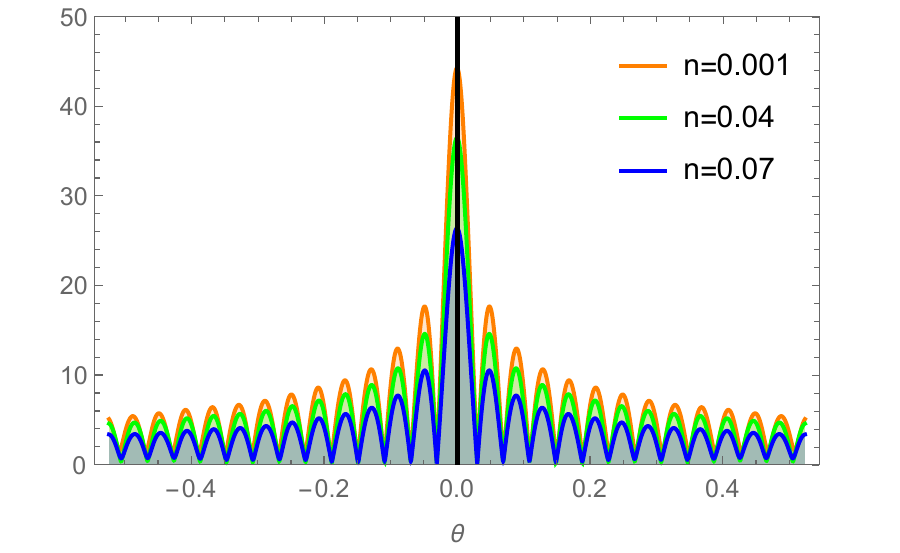}
    }
    \caption{The amplitude of $\bra \mathcal{O}\ket$ for different  $n$  with $z_h=1$ and $\omega=80$.}
    \label{m}
\end{figure}

\begin{figure}
    \centering
    \subfigure[]{
        \includegraphics[width=3in]{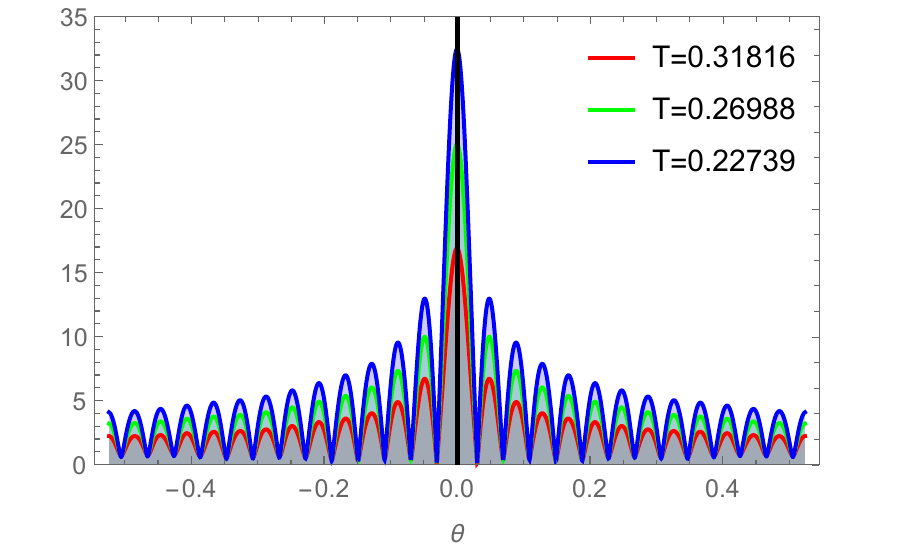}
    }
    \caption{The amplitude of $|\bra \mathcal{O}\ket|$ for different $T$  with $n=0.04$ and $\omega=80$. The  temperatures $T=0.31816,0.26988, 0.22739$ correspond to $z_h=0.7, 0.8, 0.9 $ respectively.}
    \label{n}
\end{figure}

The bulk solution is easily derived with the source given by Eq. (\ref{source})
\begin{equation}
	\text{}\mathscr{\phi}(v_e,z,\theta)=\text{}e^{-i\omega v_e}\sum_{l=0}^\infty c_{l0}\mathcal{Z}_l(z)Y_{l0}(\theta)\text{},
\end{equation}
where $\text{}\mathcal{Z}_l$ satisfies the following equation of motion
\begin{equation}
\text{}z^2F\mathcal{Z}_l''\text{}+z^2[F'+2i\omega]\mathcal{Z}_l'+[(2-2F)+zF'-z^2 l(l+1)]\mathcal{Z}_l=0\text{},\label{radial}
\end{equation}
and its asymptotic behaviour near the AdS boundary goes
\begin{equation}\label{expansion} 	 \text{}\mathcal{Z}_l=1+z\text{}\bra\mathcal{O}\ket_l +O(z^2)\text{}.
\end{equation}
And the resulting response $\text{}\bra\mathcal{O}\ket_{J_\mathcal{O}}\text{}$ is then written as
\begin{equation}\label{decomposition}\text{}\bra\mathcal{O}\ket_{J_\mathcal{O}}\text{}=e^{-i\omega v_e}\sum_{l=0}^\infty c_{l0}\bra\mathcal{O}\ket_{J_\mathcal{O}l} Y_{l0}(\theta)\text{},
\end{equation}

with
\begin{equation}
\bra\mathcal{O}\ket_{J_\mathcal{O}l}=\bra\mathcal{O}\ket_{l}+\omega. 
\end{equation}
Our main goal is to solve the radial Eq. (\ref{radial}) with
the boundary condition $\text{}\mathcal{Z}_l(0)=1\text{}$
at the AdS boundary and $F(r)$ vanishes in Eq. (\ref{radial}) at the horizon. With the help of the pseudo-spectral method~\cite{Liu:2022cev}, we obtain the corresponding numerical solution for $\text{}\mathcal{Z}_l\text{}$ and  $\text{}\mathcal{O}_l\text{}$. With the extracted $\text{}\mathcal{O}_l\text{}$, the total response can be obtained by Eq.~(\ref{decomposition}). Here we plot a typical profile of the total response $\text{}\bra\mathcal{O}\ket\text{}$ in Fig.~\ref{omega} to Fig.~\ref{n}. All the results show that the interference pattern indeed arises from the diffraction of our scalar field by  the black hole. Explicitly, Fig.~\ref{omega} shows the amplitude of $\bra \mathcal{O}\ket$ for different $\omega$ with $z_h=1$ and $n=0.04$,  we can see that the frequency $\omega$ of the Gaussian source increases the width and 
therefore reduces the wave periods, which means the total response function depends closely on the Gaussian source. 
Next we investigate the dependence of the response function $\bra\mathcal{O}\ket$ on the non-commutative strength parameter $n$.  Here we fix the parameters $z_h=1$, $\omega=80$. From Fig.~\ref{m}, we see that the amplitude of $\bra\mathcal{O}\ket$ increases with the decrease of the non-commutativity strength parameter $n$.
Similarly, we study the dependence of the total response function $\bra\mathcal{O}\ket$ on the horizon temperature $T$, which is shown in Fig.~\ref{n} with fixed $n=0.04$ and $\omega=80$. The corresponding temperatures are $T=0.31816$, $T=0.26988$ and $T=0.22739$.
Fig.~\ref{n} shows that the amplitude of $\bra\mathcal{O}\ket$ increases when the temperature $T$ decreases.

In all, the total response function depends closely on the Gaussian source and the spacetime geometry. Next, we devoted to convert this response function into an observed images with the optical system described in the section~\ref{sec1}.

\section{Holographic Einstein ring in ADS black hole}\label{sec3}

According to Eq.(\ref{transformation}), we are able to see  the observed wave on the screen which is connected with the incident wave by the Fourier transformation.  We will set  $\text{}\delta= 0.02$ for the source and $d= 0.6$ for the convex lens radius.
\begin{figure}
    \centering
    \subfigure[$\omega=11$]{
        \includegraphics[width=1.4in]{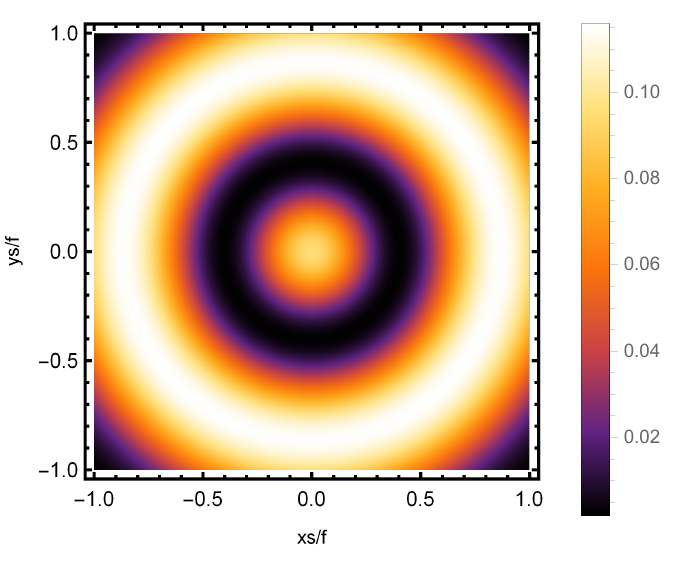}
    }
	\subfigure[$\omega=21$]{
        \includegraphics[width=1.4in]{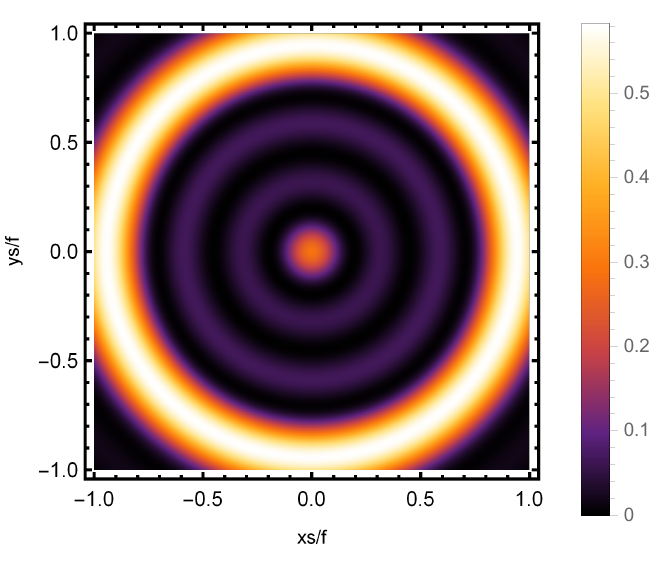}
    }
    \subfigure[$\omega=41$]{
        \includegraphics[width=1.4in]{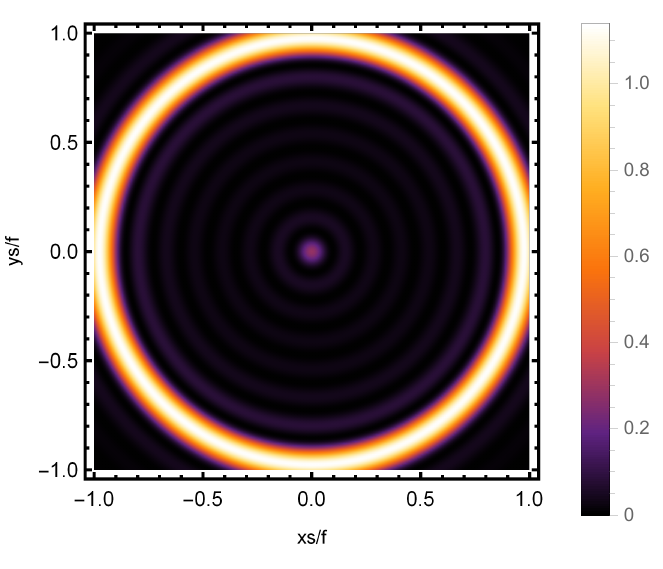}
    }
    \subfigure[$\omega=61$]{
        \includegraphics[width=1.4in]{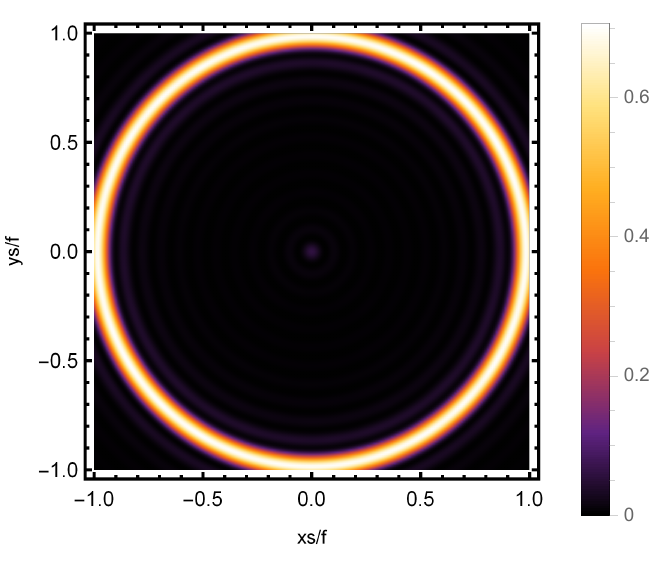}
    }
    \caption{The images of the lensed response observed at the observation angle $\theta_{obs}=0$ for different $\omega$ with $n=0.0001$ and $z_h=1$.}\label{image1}
\end{figure}

\begin{figure}
    \centering
	\subfigure[$\omega=11$]{
        \includegraphics[width=1.5in]{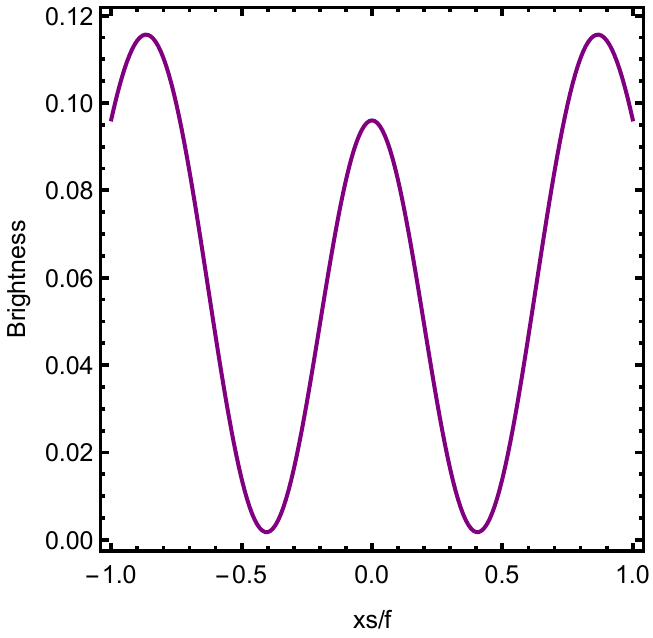}
    }
    \subfigure[$\omega=21$]{
        \includegraphics[width=1.5in]{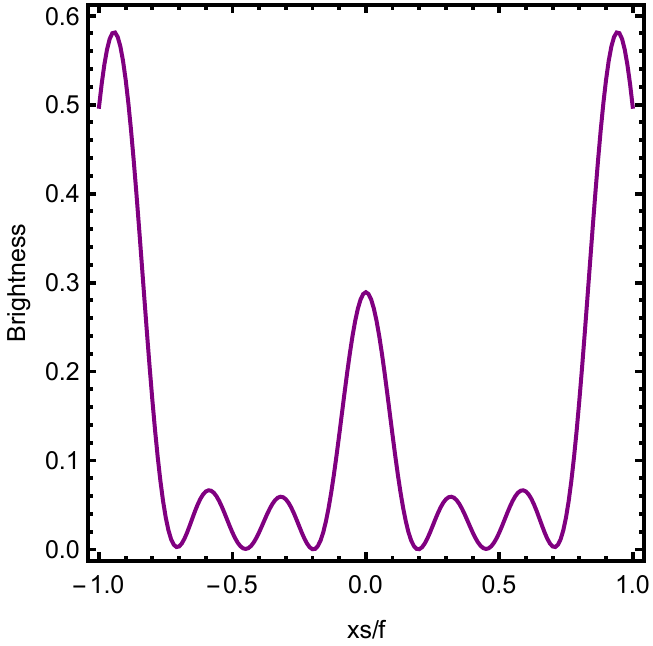}
    }
    \subfigure[$\omega=41$]{
        \includegraphics[width=1.5in]{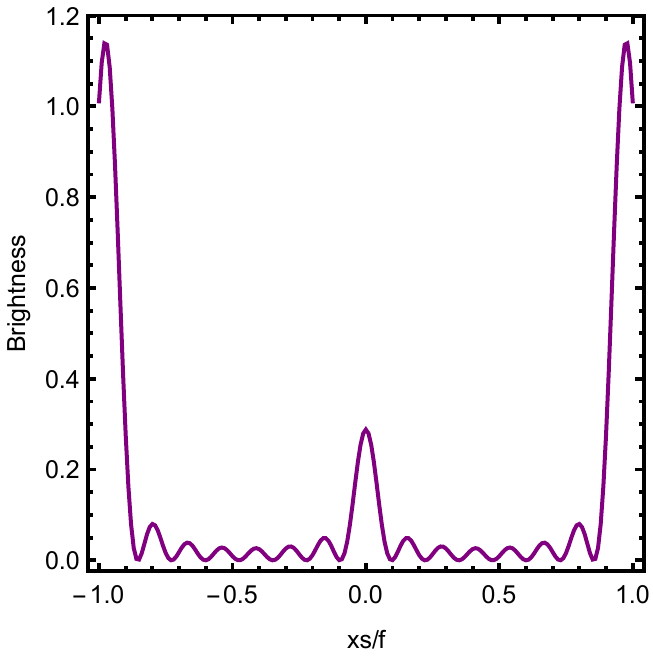}
    }
    \subfigure[$\omega=61$]{
        \includegraphics[width=1.5in]{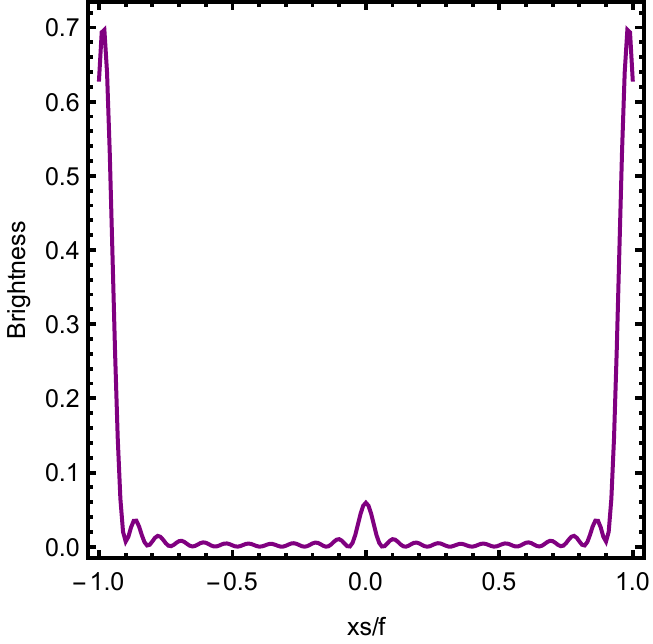}}
    \caption{The brightness of the lensed response on the screen for different $T$ with $\omega=80$ and  $n=0.0001$.}\label{image11}
\end{figure}

\begin{figure}
    \centering
    \subfigure[$n=0.0001$,$\theta_{obs}=0$]{
        \includegraphics[width=1.4in]{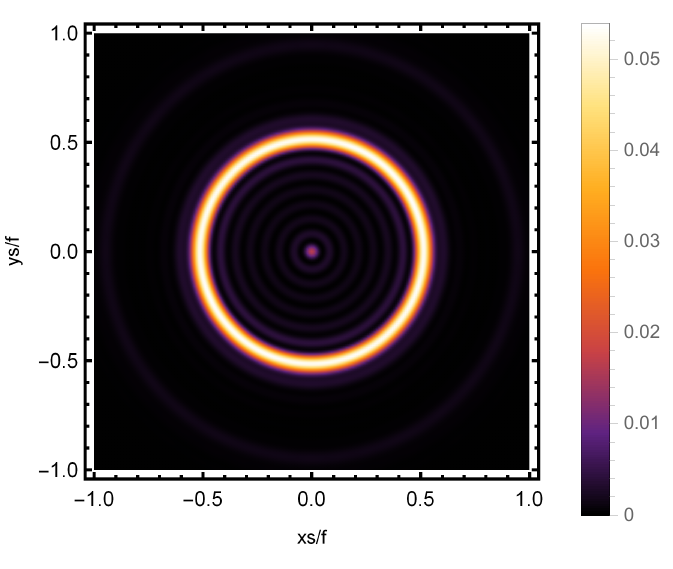}
    }
	\subfigure[$n=0.0001$,$\theta_{obs}=\frac{\pi}{6}$]{
        \includegraphics[width=1.4in]{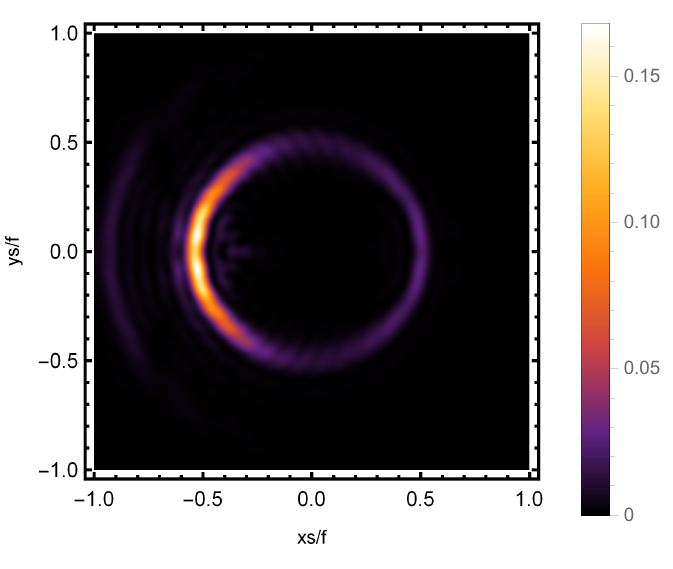}
    }
    \subfigure[$n=0.0001$,$\theta_{obs}=\frac{\pi}{3}$]{
        \includegraphics[width=1.4in]{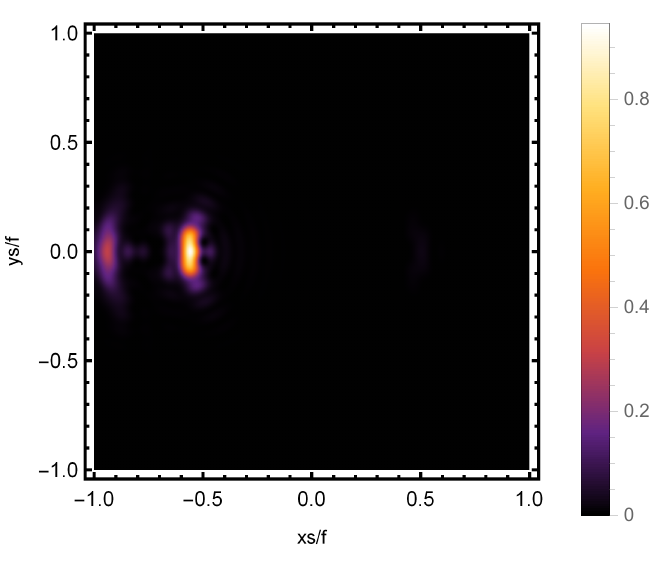}
    }
    \subfigure[$n=0.0001$,$\theta_{obs}=\frac{\pi}{2}$]{
        \includegraphics[width=1.4in]{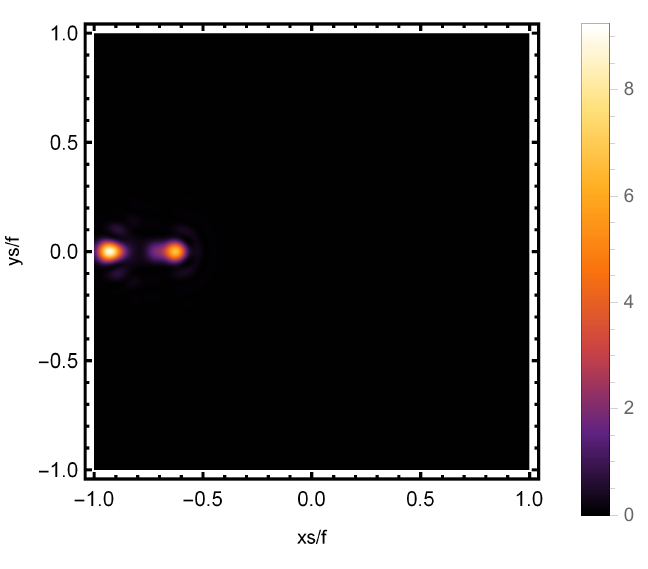}
    }
    \subfigure[$n=0.001$,$\theta_{obs}=0$]{
        \includegraphics[width=1.4in]{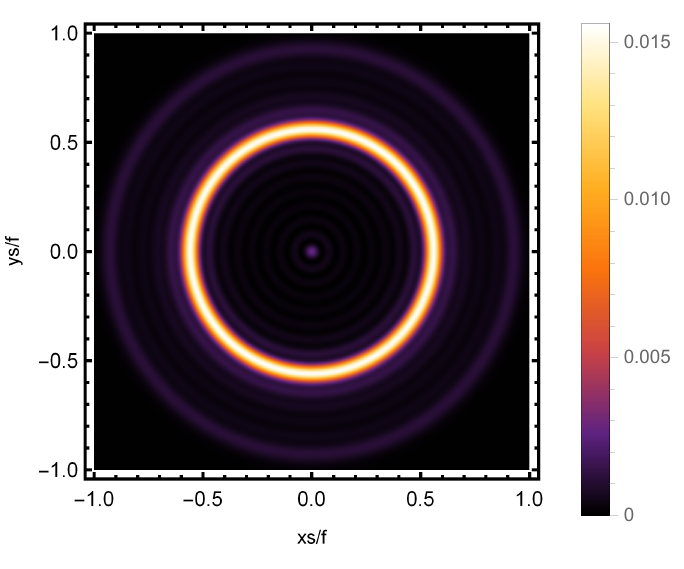}
    }
	\subfigure[$n=0.001$,$\theta_{obs}=\frac{\pi}{6}$]{
        \includegraphics[width=1.4in]{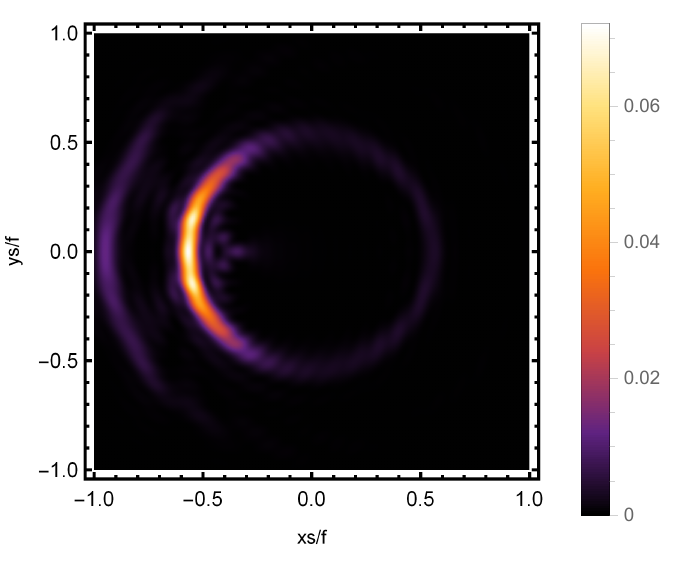}
    }
    \subfigure[$n=0.001$,$\theta_{obs}=\frac{\pi}{3}$]{
        \includegraphics[width=1.4in]{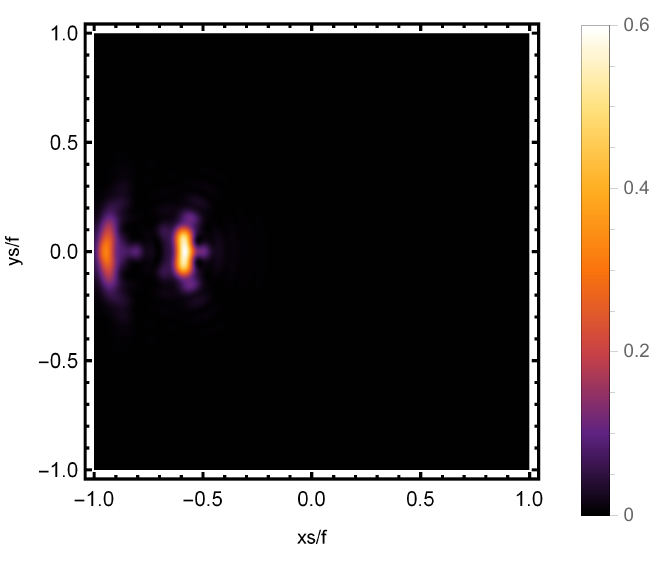}
    }
    \subfigure[$n=0.001$,$\theta_{obs}=\frac{\pi}{2}$]{
        \includegraphics[width=1.4in]{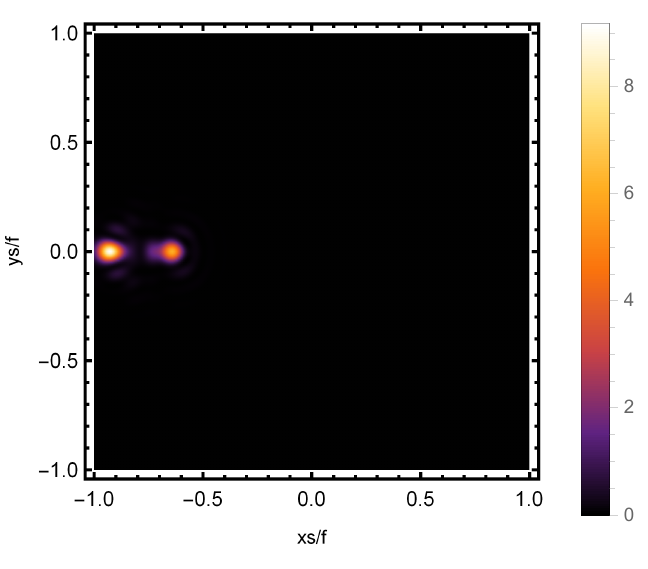}
    }
    \subfigure[$n=0.0015$,$\theta_{obs}=0$]{
        \includegraphics[width=1.4in]{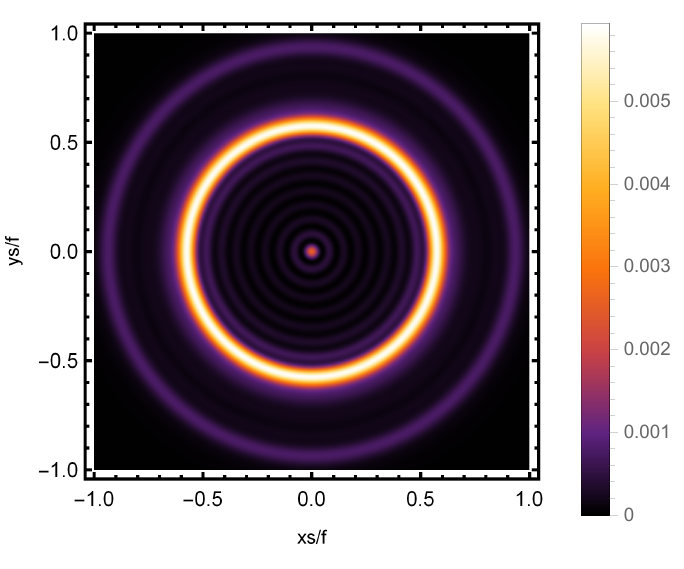}
    }
	\subfigure[$n=0.0015$,$\theta_{obs}=\frac{\pi}{6}$]{
        \includegraphics[width=1.4in]{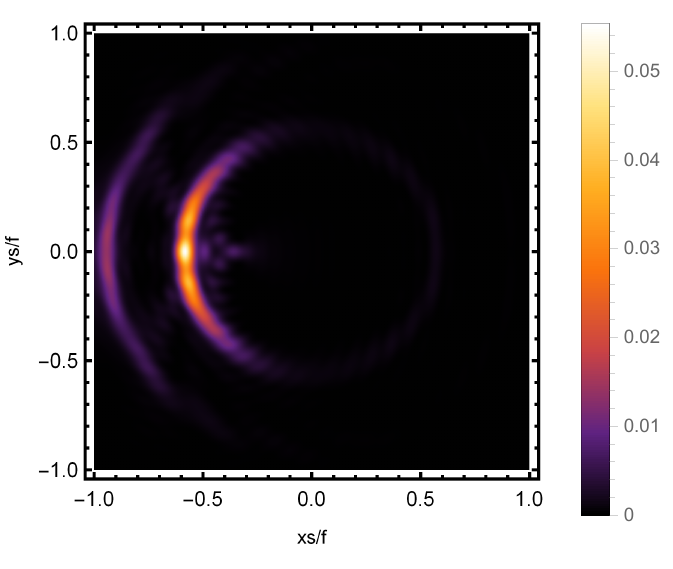}
    }
    \subfigure[$n=0.0015$,$\theta_{obs}=\frac{\pi}{3}$]{
        \includegraphics[width=1.4in]{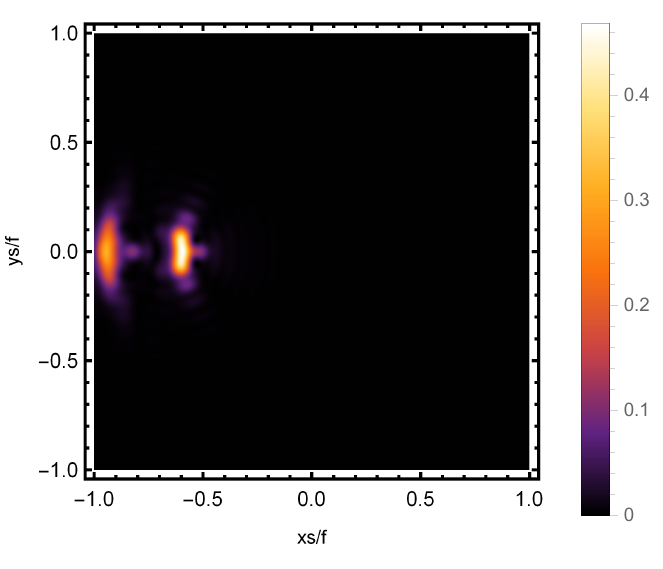}
    }
    \subfigure[$n=0.0015$,$\theta_{obs}=\frac{\pi}{2}$]{
        \includegraphics[width=1.4in]{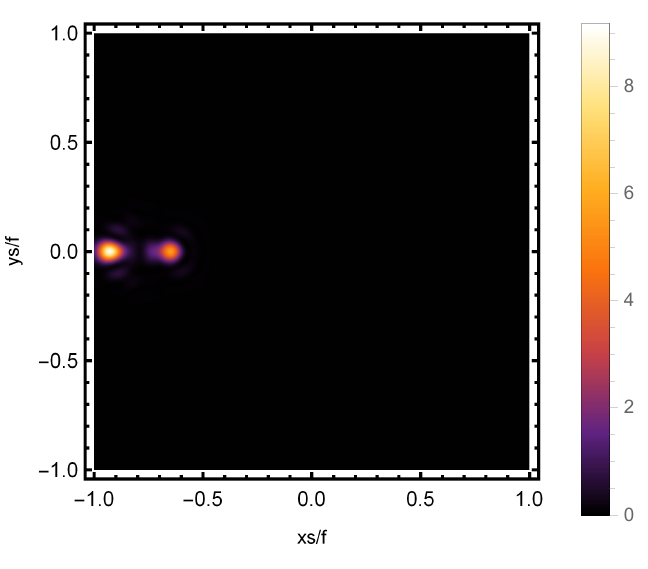}
    }
    \subfigure[$n=0.002$,$\theta_{obs}=0$]{
        \includegraphics[width=1.4in]{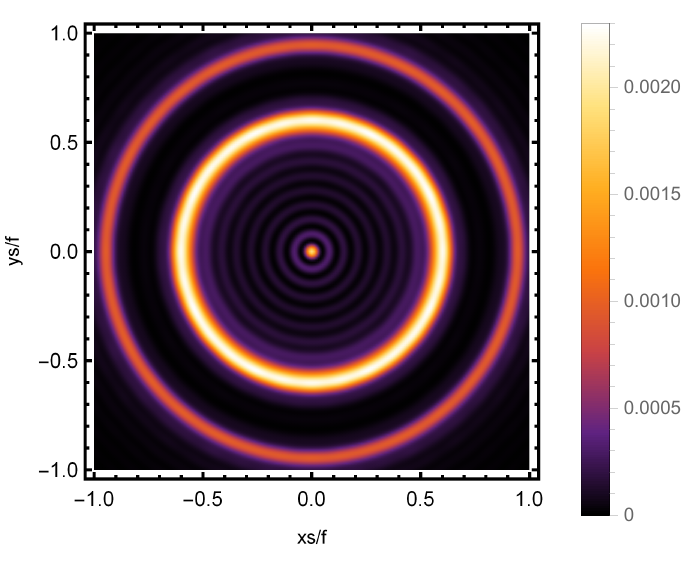}
    }
	\subfigure[$n=0.002$,$\theta_{obs}=\frac{\pi}{6}$]{
        \includegraphics[width=1.4in]{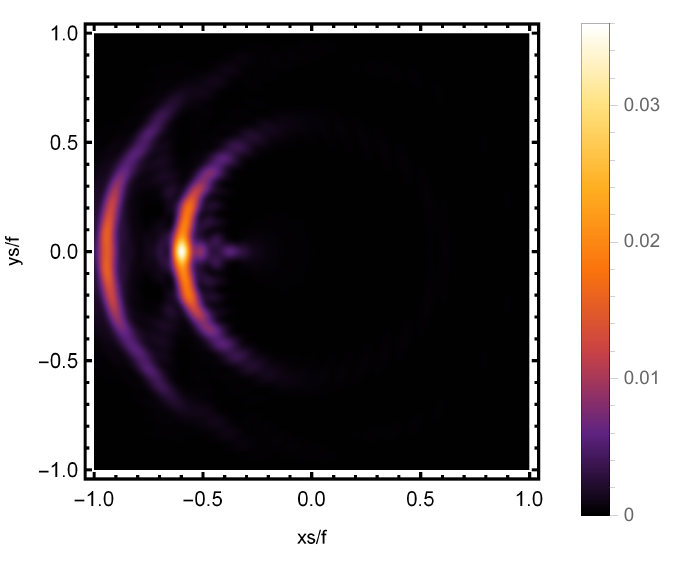}
    }
    \subfigure[$n=0.002$,$\theta_{obs}=\frac{\pi}{3}$]{
        \includegraphics[width=1.4in]{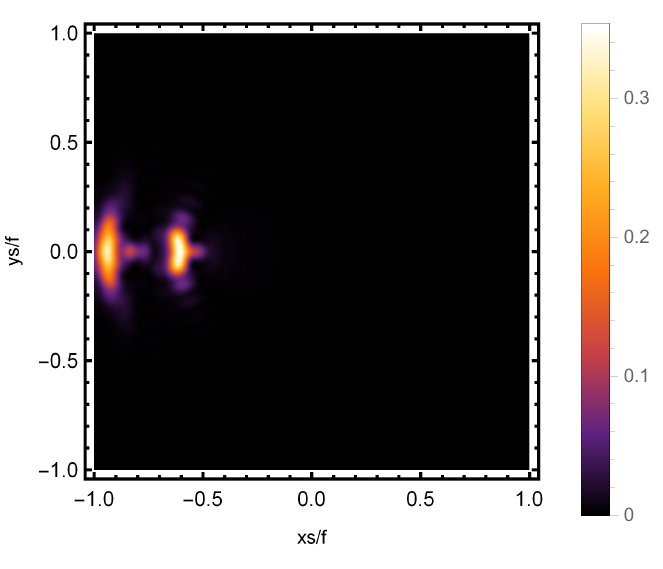}
    }
    \subfigure[$n=0.002$,$\theta_{obs}=\frac{\pi}{2}$]{
        \includegraphics[width=1.4in]{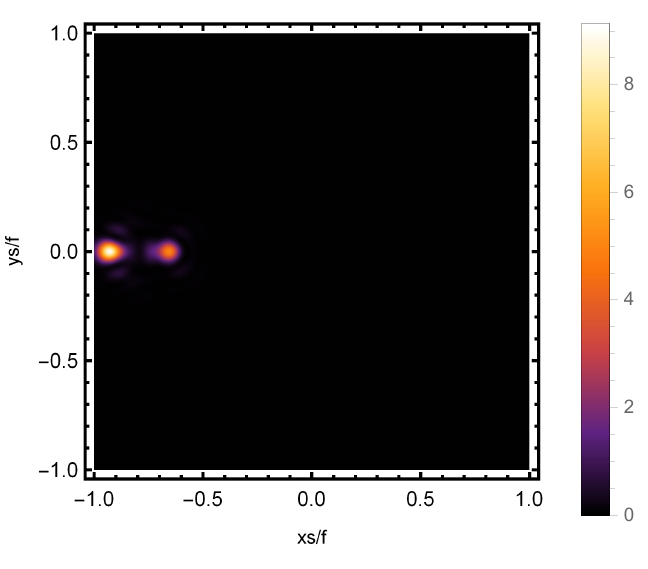}
    }
    \caption{The images of the lensed response observed at various observation angles for different $n$ with $z_h=5$ and $\omega=80$.}\label{image2}
\end{figure}

Firstly,  we consider the effect of wave source on the characteristics of the holographic Einstein image with different $\omega$ observed from the north pole with the non-commutativity strength parameter $n=0.0001$ and $z_h=1$, which is shown in Fig.~\ref{image1}. Here from the left to the right, the values of $\omega$ are $\omega=11$, $\omega=21$, $\omega=41$ and $\omega=61$, respectively. We can see that the ring is more clear for the high frequency.  The corresponding curves which show the brightness of the lensed response on the screen for the same parameters are shown in Fig.~\ref{image11}. The peaks of the curves correspond to the 
radius of the Einstein ring. From ths figure, we find 
the higher the frequency becomes, the sharper the Einstein ring becomes. This makes sense because the image can be well captured by the geometric optics approximation in the high frequency limit.

\begin{figure}
    \centering
    \subfigure[$n=0.0001$]{
        \includegraphics[width=1.45in]{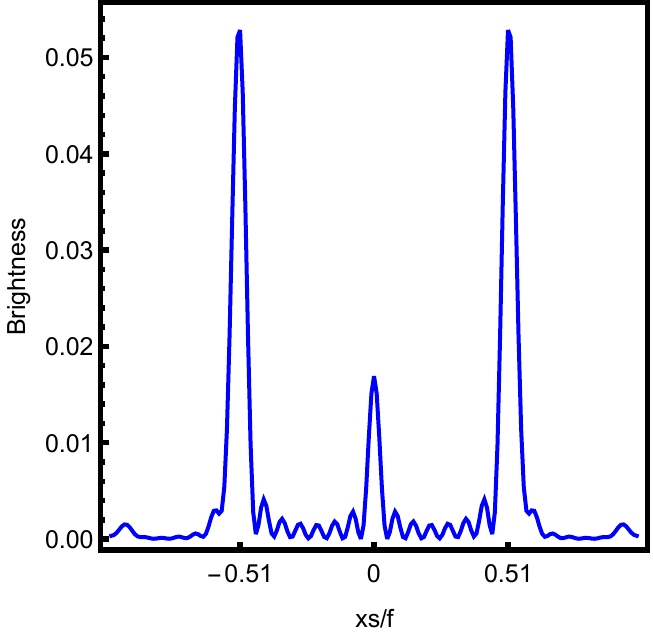}
    }
	\subfigure[$n=0.001$]{
        \includegraphics[width=1.45in]{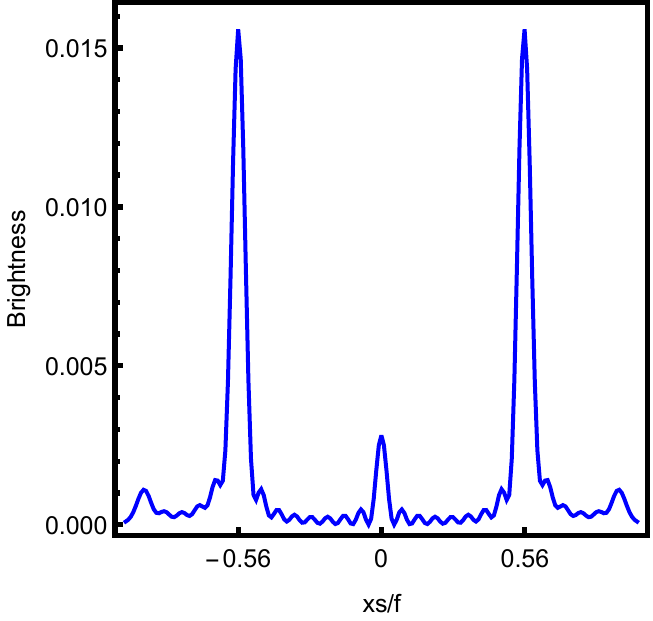}
    }
    \subfigure[$n=0.0015$]{
        \includegraphics[width=1.45in]{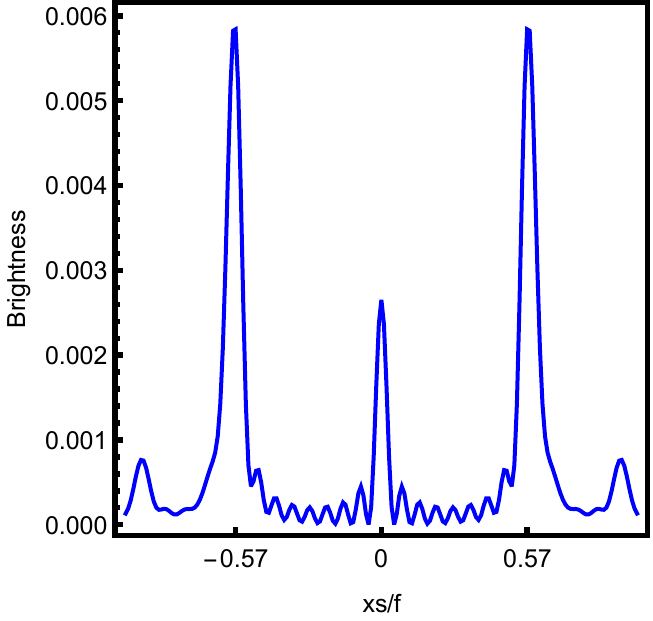}
    }
     \subfigure[$n=0.002$]{
        \includegraphics[width=1.45in]{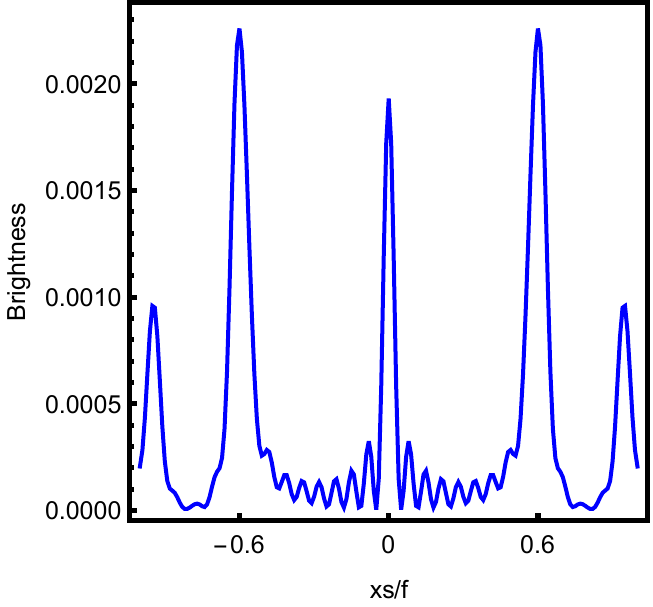}
    }
    \caption{ The brightness of the lensed response on the screen for different $n$ with $\omega=80$ and  $z_h=1$.}\label{image21}
\end{figure}

Next, we discuss the influence of different non-commutativity strength parameter $n$ on the Einstein ring shown in Fig.~\ref{image2}. Supposing the observer is located at different positions of AdS boundary with the change of the non-commutativity strength parameter $n$ for $\text{}z_h=5$ and frequency $\text{}\omega = 80$. When the observer is located at the position $\text{}\theta_{obs}=0$, which means the observation is located at the north pole of the AdS boundary. A series of axisymmetric concentric rings appear in the image, and one of them is particularly bright which is shown in the left-most column of Fig.~\ref{image2}. Explicitly from top to bottom, the noncommutativity strength parameter $n$ increases from $n=0.0001$ to $n=0.002$. And more, as the parameter $n$ increases, the brightest ring is away from the center. All these can be clearly seen in Fig.~\ref{image21} which also shows that the brightness peak of lensed response is far away from the center as the parameter $n$ increases for the same parameter. Next we fix the observed position to $\text{}\theta_{obs}=\pi/6$ (the second column from the left shown in Fig.~\ref{image2}). We see the light arcs instead of a strict axisymmetric ring, which are consistent with ~\cite{Zeng:2023tjb,Zeng:2023zlf}. And the positions of the light arcs are away from the center as the parameter $n$ increases. That is to say, a series of axisymmetric concentric rings still exist in the image. And from the top to the bottom, we see that the brightness of the ring decreases when the parameter $n$ increases. However, when we move to $\theta_{obs}=\pi/3$, we see the light arcs become much smaller. When the observer is at $\theta_{obs}=\pi/2$, all left are bright spots shown on the right-most column of Fig.~\ref{image2}. And as the parameter $n$ increases, the bright spot becomes far away from the center.

\begin{figure}
    \centering
	\subfigure[$T=0.299$]{
        \includegraphics[width=1.4in]{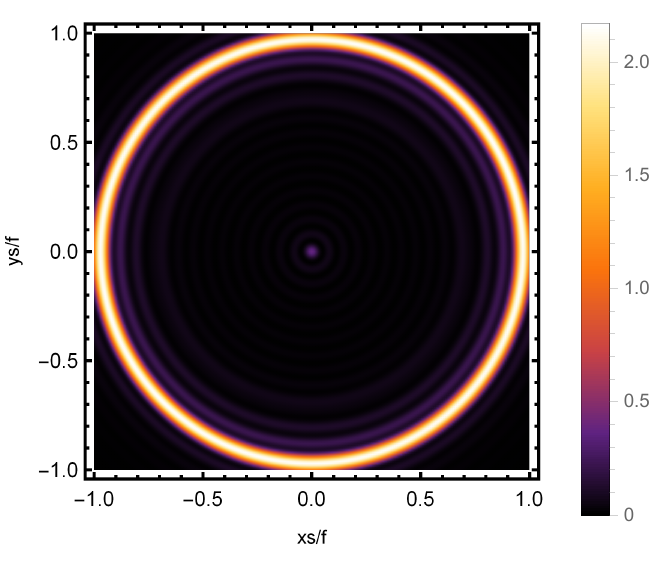}
    }
    \subfigure[$T=0.393$]{
        \includegraphics[width=1.4in]{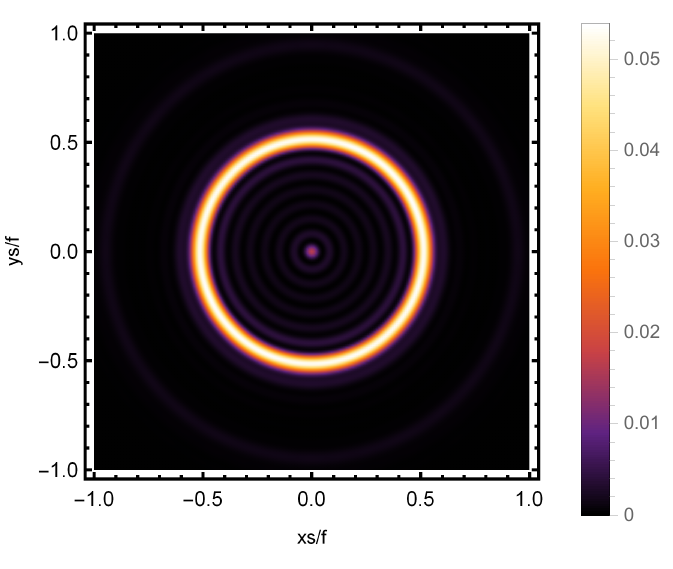}
    }
    \subfigure[$T=0.485$]{
        \includegraphics[width=1.4in]{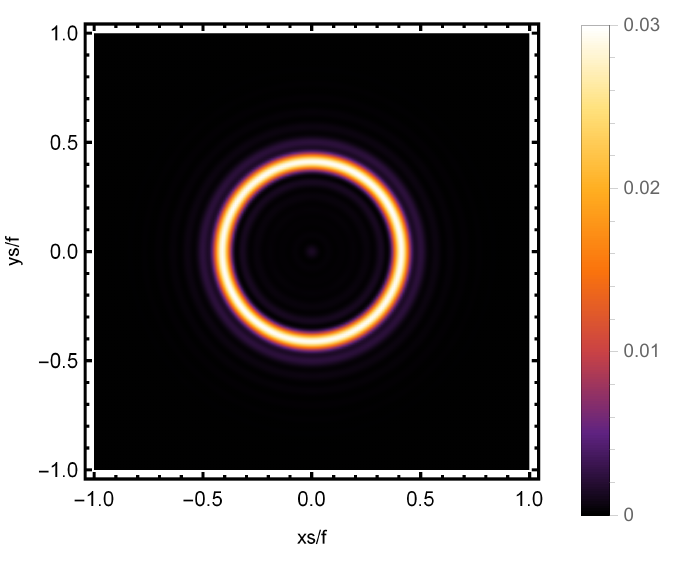}
    } \subfigure[$T=0.558$]{
        \includegraphics[width=1.4in]{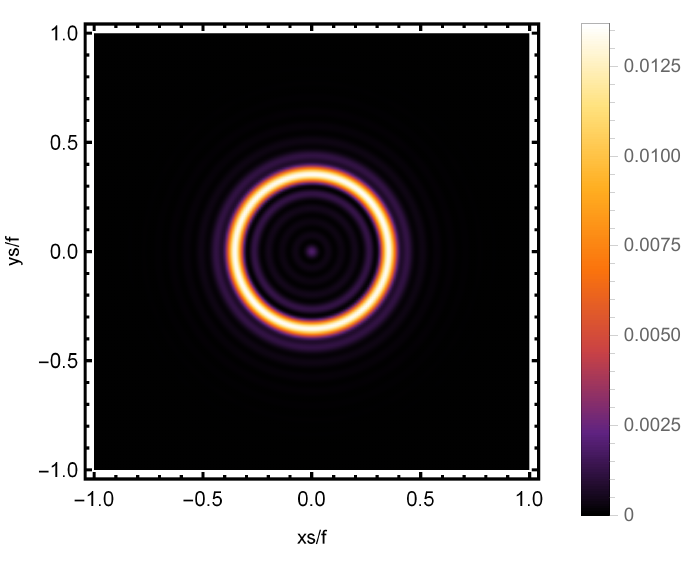}
    }
    \caption{The images of the lensed response observed at the observation angle $\theta_{obs}=0$  for different $T$ with $n=0.0001$ and $\omega=80$. From (a) to (d), the horizons correspond  to $z_h=3,5,7,9$ respectively.  }\label{sharpimage_2}
\end{figure}

\begin{figure}
    \centering
	\subfigure[$T=0.299$]{
        \includegraphics[width=1.45in]{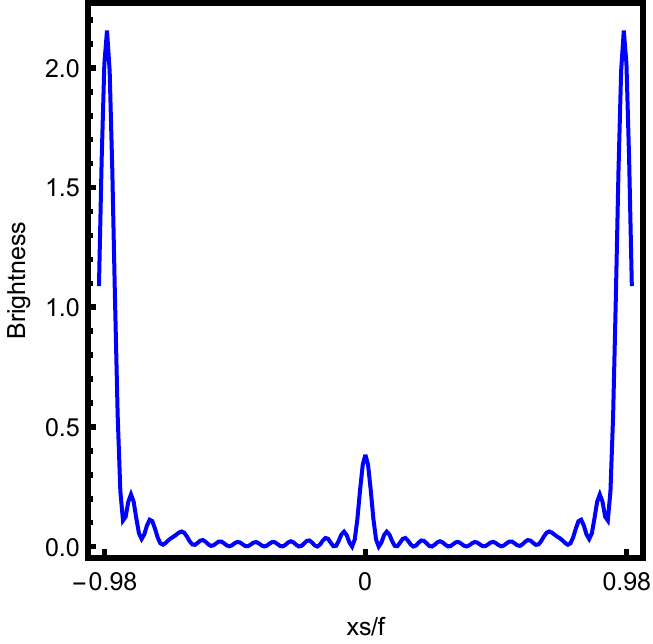}
    }
    \subfigure[$T=0.393$]{
        \includegraphics[width=1.45in]{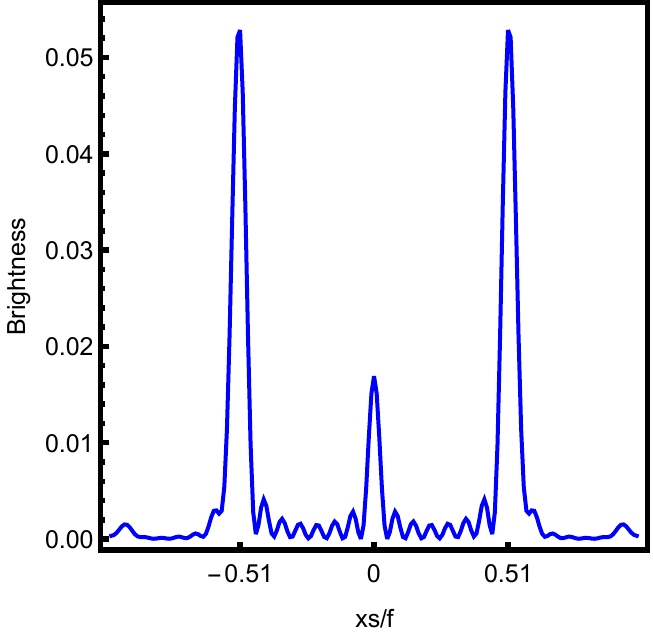}
    }
    \subfigure[$T=0.485$]{
        \includegraphics[width=1.45in]{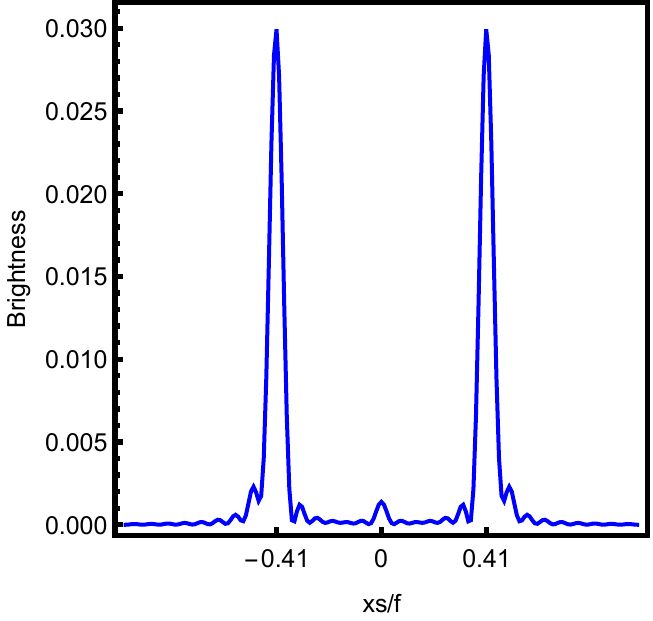}
    }
    \subfigure[$T=0.558$]{
        \includegraphics[width=1.45in]{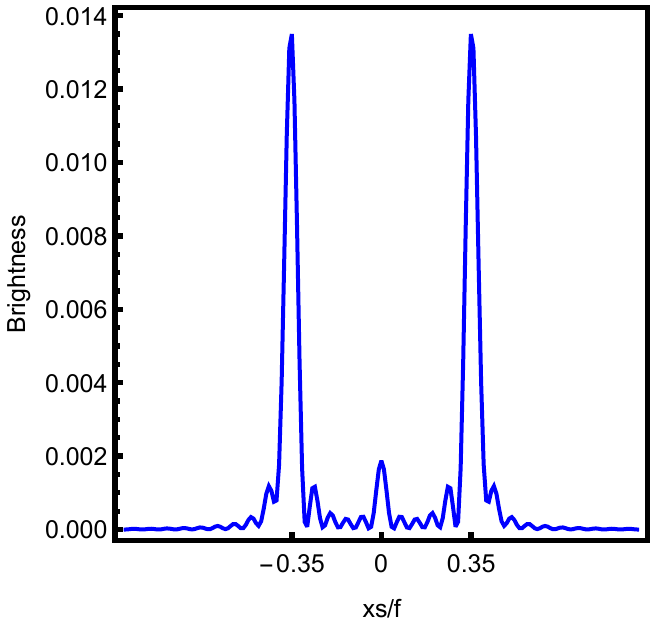}
    }
    \caption{The brightness of the lensed response on the screen for different $T$ with $\omega=80$ and $n=0.0001$. From (a) to (d), the horizons correspond  to $z_h=3,5,7,9$ respectively.}\label{sharpimage_221}
\end{figure}

For better understanding the above holographic Einstein image, we also study the impact of the horizon temperature on the lensed response observed with the fixed observation angle $\text{}\theta_{obs}=0$. We fix noncommutativity strength parameter $\text{}n=0.0001$ and frequency $\text{}\omega=80$ shown in Fig.~\ref{sharpimage_2} and the corresponding brightness of the lensed response is in Fig.~\ref{sharpimage_221}. The four subfigures corresponds to $T=0.299$, $T=0.393$, $T=0.485$ and $T=0.558$.
We find that the brightest ring moves to the center as the temperature increases.  This conclusion also can be obtained from Fig.~\ref{temperature_2}, in which the brightness for different temperatures are shown.


\section{The comparison between the holographic results and optical results}\label{sec4}
In this section, we compare the results from the holographic dual with the results from the geometrical optics. At the position of the photon sphere, there exists a brightest ring in the image. In this section, we calculate this
brightest ring  from the viewpoint of optical geometry. In a spacetime with metric in Eq.(\ref{EQ2.3}), the ingoing angle of photons from boundary is expressed with the conserved energy $\omega$ and the angular momentum $\mathcal{L}$. For generality, we choose the coordinate system in order to let the photon orbit lying on the equatorial plane $\theta=\pi/2$.
The 4-vector $\text{} u^a=(d/d\nu)^a\text{}$ satisfies ~\cite{
Hashimoto:2018okj,Zeng:2020vsj}
\begin{equation}
\text{}-F(r)\left(\diff{t}{\nu}\right)^2\text{} +\frac{1}{F(r)}\left(\diff{r}{\nu}\right)^2 +r^2\sin^2\theta \left(\diff{\mathscr{\phi}}{\nu}\right)^2=0\text{},
\end{equation}
or equivalently,
\begin{equation}
	\text{}\dot r^2\text{}=\omega^2-\mathcal{L}^2\mathcal{R}\text{},
\end{equation}
where $\text{}\mathcal{R}=F(r)/r^2\text{}$, $\text{}\omega =F(r)\dot{t}\text{}$, $\text{}\mathcal{L}=r^2 \dot{\phi}\text{}$, and $\text{}\dot r \notice \frac{\partial r}{\partial \nu}\text{}$, $\text{}\dot t\notice \frac{\partial t} {\partial \nu}\text{}$, $\text{}\dot \phi \notice \frac{\partial \phi}{\partial \nu}\text{}$.

The ingoing angle $\theta_\text{in}$ with normal vector of boundary $n^b=\pd/\pd{r}^b$ should be ~\cite{
Hashimoto:2018okj,Hashimoto:2019jmw}
\begin{eqnarray}\text{}\cos\theta_\text{in}&=&\text{}\frac{g_{jk}u^j n^k}{|u||n|}\when_{r=\infty}\nonumber\\
&=&\text{}\sqrt{\frac{\dot r^2/F}{\dot {r}^2/F+\mathcal{L}^2/r^2}}\when_{r=\infty}\text{},
\end{eqnarray}
and this implies
\begin{eqnarray}
	\text{}\sin^2\theta_\text{in}&=&1-\cos^2\theta_\text{in}\nonumber\\
 &=&\frac{\mathcal{L}^2 \mathcal{R}}{\dot r^2+\mathcal{L}^2\mathcal{R}} \when_{r=\infty}\nonumber\\
 &=&\frac{\mathcal{L}^2}{\omega^2}\text{}.
\end{eqnarray}

\begin{figure}[ht]
	\centering
	\includegraphics[trim=0.4cm 0.4cm 3.4cm 2.8cm, clip=true, scale=0.8]{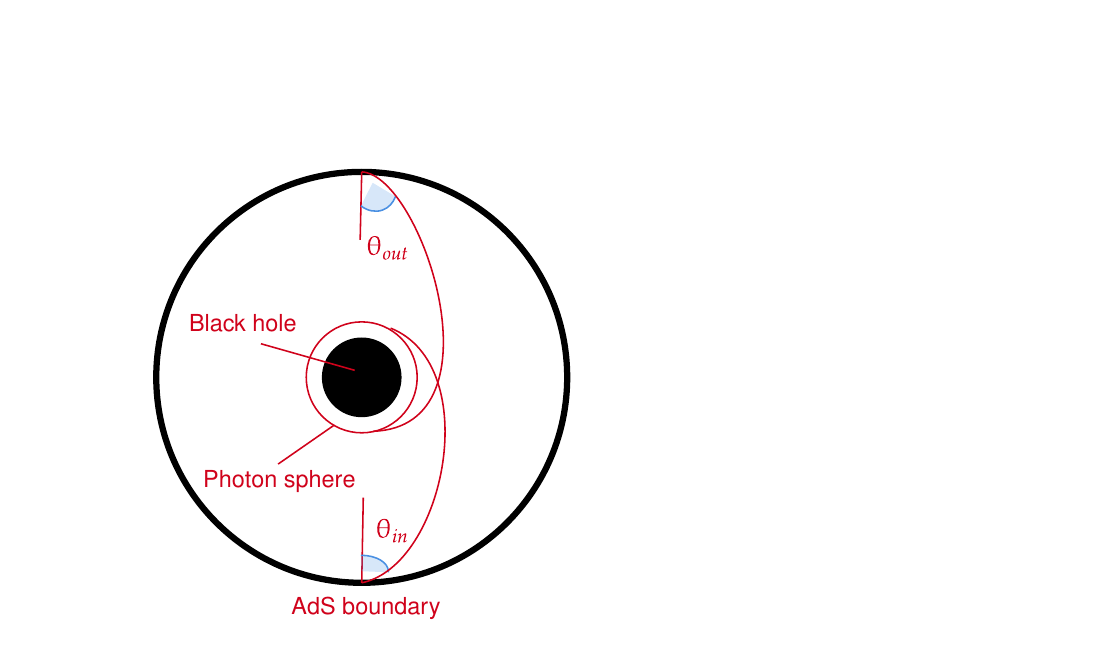}
	\caption{ The ingoing angle $\theta_{in}$ and outgoing $\theta_{out}$  angle  of the photon at the photon sphere. }\label{reach_and_response2}
\end{figure}

The corresponding ingoing angle $\text{}\theta_\text{in}\text{}$ of photon orbit from boundary satisfies that
\begin{equation}	\text{}\sin\theta_{in}=\text{}\frac{\mathcal{L}}{\omega}\text{},
\end{equation}
which is shown in Fig.~\ref{reach_and_response2}. The above relation is still valid when the light is located at the photon sphere. Label the angular momentum as $L_s$, which is determined by the following conditions
\begin{eqnarray}
\text{}\dot{r}=0\text{},   \ \ \   \text{}\frac{d\mathcal{R}}{dr}=0\text{}.
\end{eqnarray}

In the geometrical optics, the angle $\theta_{in}$ gives the angular distance of the image of the incident ray from the zenith if an observer is located on the AdS boundary looks up into the AdS bulk. When two end points of the geodesic and the center of the black hole are in alignment, the observer sees a ring with a radius corresponding to the incident angle $\theta_{in}$ because of axisymmetry~\cite{Hashimoto:2018okj}.
\begin{figure}[ht]
	\centering
	\includegraphics[trim=0.4cm 1.8cm 0.4cm 0.8cm, clip=true, scale=0.8]{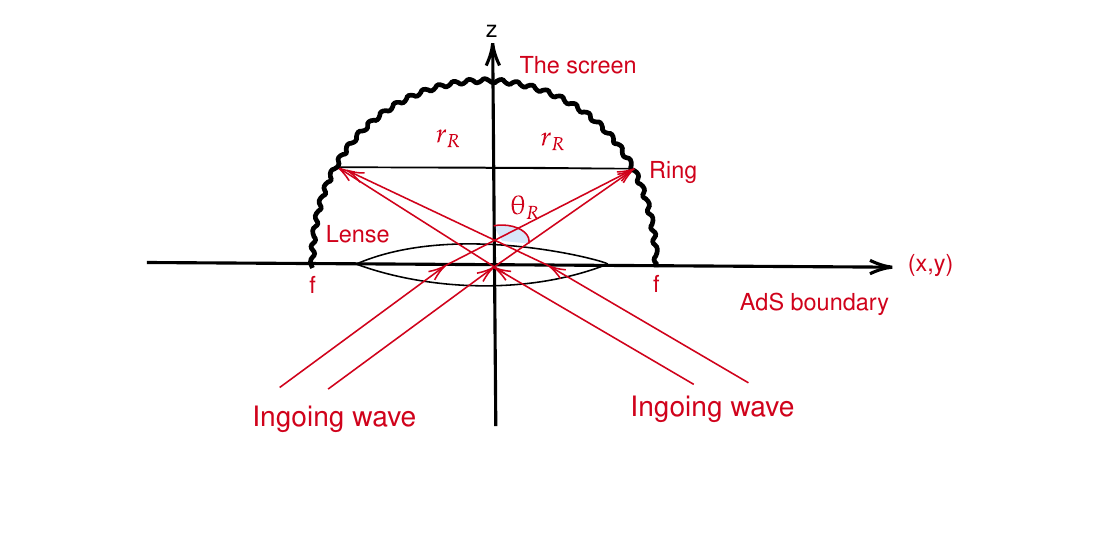}
	\caption{ The relation between the $\text{}\theta_R $ and $\text{}r_R$. }\label{irreg1_1}
\end{figure}

\begin{figure}
    \centering
    \subfigure[$n=0.0001$]{
        \includegraphics[width=2in]{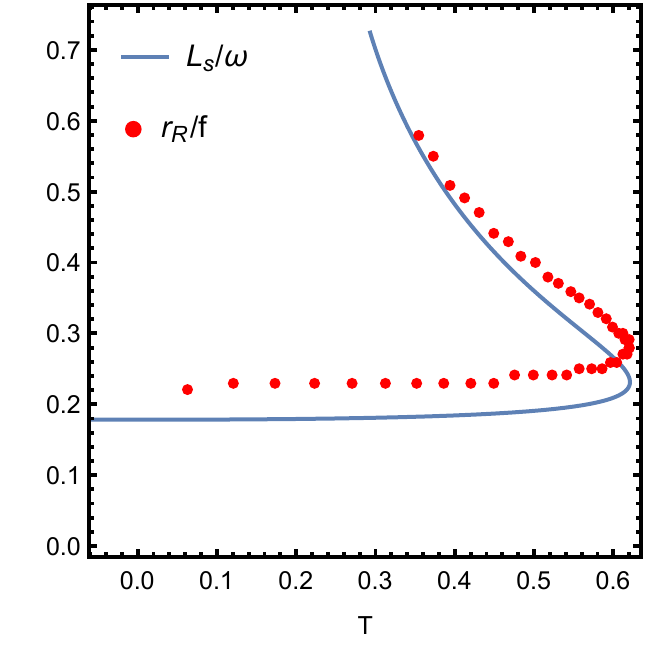}
    }
	\subfigure[$n=0.001$]{
        \includegraphics[width=2in]{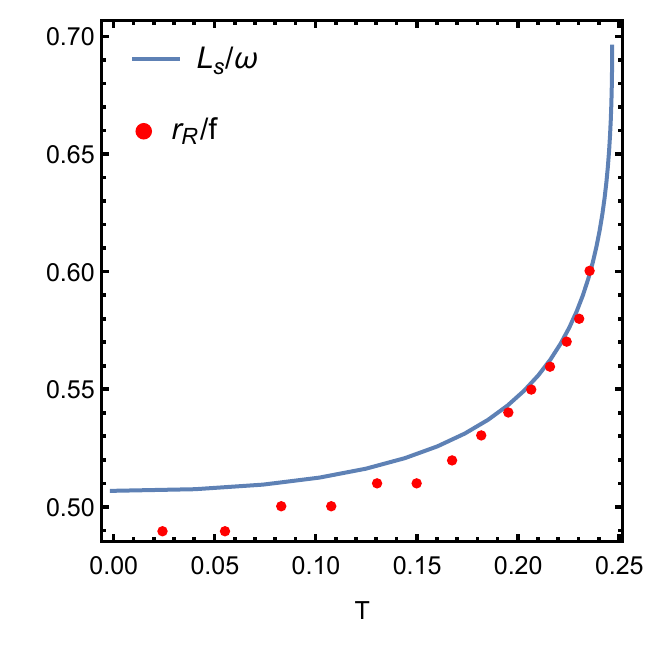}
    }
    \caption{Comparison between  the results obtained by wave optics and geometric optics for different $ n$ with $\omega=80$.}\label{sharpimage1_1}
\end{figure}

\begin{figure}
    \centering
    \subfigure[$\omega=80$]{
        \includegraphics[width=2in]{c0.pdf}
    }
	\subfigure[$\omega=120$]{
        \includegraphics[width=2in]{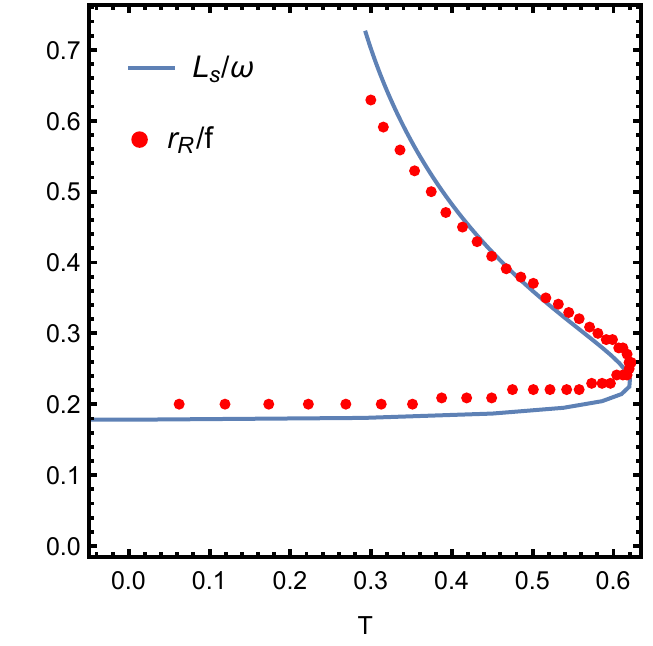}
    }
    \subfigure[$\omega=160$]{
        \includegraphics[width=2in]{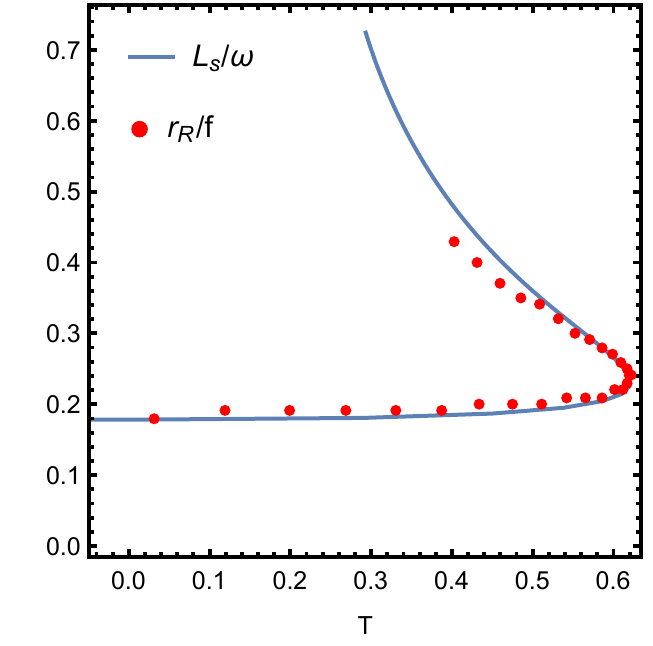}
    }
    \caption{Comparison between  the results obtained by wave optics and geometric optics for different $\omega$ with $n=0.0001$.}\label{sharpimage2_1}
\end{figure}

In addition, with Fig.~\ref{irreg1_1}, we can
obtain the angle of the Einstein ring, which is  
\begin{equation}
\sin\theta_{R}=\frac{r_R}{f}.
\end{equation}
According to \cite{Hashimoto:2018okj}, for a sufficiently large $l$, we have $\sin\theta_{R}=\sin\theta_{in}$. Then the desired relation is  
\begin{equation}
\text{}\frac{r_{R}}{f}=\text{}\frac{L_s}{\omega}\text{}, 
\end{equation}
here $L_s$ is the angular momentum at the photon sphere. This relation   also  can be proved numerically. For different $n$, we find $\frac{r_R}{f}$ fits well with  $\frac{L_s}{\omega}$, which is   shown in Fig.~\ref{sharpimage1_1}. Note that for the large $n$, the fitting  is more difficult. Even for small $n$,  $\frac{r_R}{f}$  and  $\frac{L_s}{\omega}$ are not fully the same. In Fig.~\ref{sharpimage2_1}, we discuss the effect of the frequency on the fitting. It is obvious that the larger the frequency, the preciser the fitting.

\section{Conclusions}
\label{sec5}

In the framework of AdS/CFT correspondence,  we  studied the holographic Einstein images of a non-commutative Schwarzschild black hole. We considered a $(2+1)$ dimensional boundary conformal field theory on a 2-sphere $S^2$ at a finite temperature. Under a time-dependent localized Gaussian source $J_{\mathcal{O}}$ with the frequency $\omega$, we derived the local response function. Explicitly, the absolute amplitude $|\bra \mathcal{O}\ket|$ increases with the decrease of the non-commutative parameter $n$ and  increases with the decrease of the temperature $T$. We also investigated the effect of the frequency of the wave source on the 
 response function, and found the period  decreases as the frequency increases.

With a virtual optical system with a convex lens, we further  obtained the Einstein rings. We found  the holographic ring always appears with the concentric stripe when the observer located at the north pole while it changes into a luminosity-deformed ring, or bright spot as the observer away from the north pole. The effects of the noncommutative parameter $n$ and temperature on the ring were investigated too. It was found that   the ring radius becomes  larger as  the parameter increases,   and it becomes smaller as the   temperature increases. 

To confirm our results, we also investigated the Einstein ring via geometric optics. Especially, we obtained the ingoing angle of the photon. We found that at the Einstein ring, the ingoing  angle is the same nearly as that obtained via holography. In addition, we found that frequency affects the fitting accuracy. The higher the frequency, the   preciser the result.

\section*{Acknowledgements}{This work is supported  by the National
Natural Science Foundation of China (Grants Nos. 11675140, 11705005,   12375043), Innovation and Development Joint  Foundation of Chongqing Natural Science  Foundation (Grant No. CSTB2022NSCQ-LZX0021) }, and Basic Research Project of Science and Technology Committee of Chongqing (Grant No. CSTB2023NSCQ-MSX0324).

\end{document}